\documentclass[reprint,amsmath,amssymb,aps,superscriptaddress]{revtex4-1}
\usepackage{graphicx}
\usepackage{dcolumn}
\usepackage{bm}
\usepackage{braket}
\usepackage{color}
\usepackage{pstricks,pst-grad,color}
\usepackage{appendix}
\usepackage{here}
\usepackage{url}
\usepackage{hyperref}
\usepackage{autobreak}

\usepackage{comment}

\newcommand{\del}{\partial}

\newcommand{\cebipd}{\mathrm{Ce_{3}Bi_{4}Pd_{3}}}
\usepackage{tikz}
\usetikzlibrary{shapes,calc}
\usetikzlibrary{shapes.geometric,arrows.meta,decorations.markings}

\begin{document}

\title{Effects of strong correlations on the nonlinear response in Weyl-Kondo semimetals}
\author{Akira Kofuji}
\email{kofuji.akira.46c@st.kyoto-u.ac.jp}
 \affiliation{Department of Physics, Kyoto University, Kyoto 606-8502, Japan}
\author{Yoshihiro Michishita}
\email{michishita.yoshihiro.56e@st.kyoto-u.ac.jp}
 \affiliation{Department of Physics, Kyoto University, Kyoto 606-8502, Japan}
\author{Robert Peters}%
 \email{peters@scphys.kyoto-u.ac.jp}
 \affiliation{Department of Physics, Kyoto University, Kyoto 606-8502, Japan}
\date{\today}

\begin{abstract}
Nonlinear responses give rise to various exciting phenomena, which are forbidden in linear responses. Among them, one of the most fascinating phenomena is the recently observed giant spontaneous Hall effect in $\cebipd$.
This material is a promising candidate for a Weyl-Kondo semimetal, and this experiment implies that strong correlation effects can enhance the nonlinear Hall effect.
However, most theoretical studies on nonlinear responses have been limited to free systems, and the connection between nonlinear responses and strong correlation effects is poorly understood. 
Motivated by these experiments and recent theoretical advances to analyze strong correlation effects on the nonlinear response, we study a periodic Anderson model describing $\cebipd$ using the dynamical mean-field theory. We calculate the nonlinear longitudinal conductivity and the nonlinear Hall conductivity using the Kubo formula extended to the nonlinear response regime and clarify their temperature dependences. We numerically show that strong correlations can enhance nonlinear conductivities, and we conclude that the magnitude of the experimentally observed giant nonlinear Hall effect can be explained by strong correlation effects.
\end{abstract}

\maketitle

\section{Introduction}
Onsager's reciprocal theorem imposes severe restrictions on linear responses in time-reversal symmetric systems. For example, the Hall effect is forbidden in this situation\cite{onsager_1931,kubo_1957}.
On the other hand, nonlinear responses are free from such restrictions.
The inversion symmetry breaking is a crucial ingredient for nonlinear responses, which leads to a variety of remarkable transport phenomena, such as the nonlinear Hall effect\cite{Sodemann_2015,Du_2018,Zhou_2020,You_2018,Ma_2019_hall,Kang_2019}, non-reciprocal current\cite{Rikken_1997,Rikken_2001,Morimoto_2018,tokura_review_2018}, shift current\cite{baltz_1981,kral_2000,sipe_2000,auston_1972,glass_1974,tan_2016}, and injection current\cite{laman_1999,sipe_2000}.
Moreover, it has been shown that the band topology is closely related to linear responses and nonlinear responses.
Remarkably, the nonlinear Hall effect is closely connected with the "Berry curvature dipole," which is the integrated gradient of the Berry curvature in momentum space\cite{Sodemann_2015}.

Because of its relationship to the Berry curvature, the nonlinear Hall effect can be an excellent tool to investigate the topological properties of materials, which are difficult to study with conventional experimental methods.
For example, in strongly correlated topological materials, such as topological Kondo insulators\cite{dzero_2010,kim_2014,hagiwara_2016,dzero_2016} and Weyl-Kondo semimetals\cite{Xu_2017,qimiao_si_2018,grefe_prb_2020,Sarah_2021,dzsaber_weyl_kondo_2017,Guo_2018,Dzsaber_2021,Kushwaha_2019,dzsaber_2019,Fuhrman_2020}, quasiparticle bands near the Fermi surface are strongly renormalized and become extremely narrow \cite{PhysRevLett.114.177202,PhysRevB.93.235159}.
Therefore, direct observation with angle-resolved photoemission spectroscopy(ARPES) of the topological surface states and their spin-texture, which would be a good indication for the topological nature of the martial, is still very challenging.
On the other hand, the nonlinear Hall effect can be an alternative method to study topological materials and especially to study  strongly correlated topological materials using transport measurements that can be considered relatively easy.

For example, Weyl-Kondo semimetals are heavy fermion materials hosting Weyl nodes driven by the Kondo effect. On the one hand, Weyl semimetals have been found to exhibit various exciting phenomena, such as the chiral anomaly-induced magnetochiral anisotropy\cite{kharzeev_2014,Li_2016} and the bulk spin Hall effect\cite{sun_2016,lonchakov_2019}. On the other hand, heavy fermion materials are the prototypical strongly correlated materials showing various correlation-induced phases, such as antiferromagnetism\cite{si_2010,kjems_1988,coleman_2001,coleman_2007} and non-fermi liquid behavior\cite{coleman_2001,von_1996,coleman_2007}. Thus, Weyl-Kondo semimetals are anticipated to be a platform to elucidate the interplay between nontrivial topology and strong correlations. 

Based on these considerations,
the Hall effect was measured in $\cebipd$\cite{Dzsaber_2021}, which is a promising candidate for a Weyl-Kondo semimetal. Surprisingly, even without time-reversal symmetry breaking, a giant spontaneous Hall effect has been observed that has a Hall angle $10^{3}$ times larger than that of a weakly interacting Weyl semimetal calculated $\it{ab}$ $\it{initio}$\cite{Zhang_2018}.
This result is not only an indirect evidence that $\cebipd$ is a Weyl-Kondo semimetal but also implies that the nonlinear Hall effect is enhanced by strong correlation effects.

This experiment suggests that strong correlations have a significant influence on nonlinear responses. However, most theoretical studies on nonlinear responses have been limited to semiclassical discussions and analyses of free/noninteracting systems\cite{sipe_2000,Sodemann_2015,Morimoto_2018}. Intuitively, one might think that strong correlations in Kondo materials hinder transport, as quasiparticle bands become heavy. On the other hand, strong correlations can enhance the density of states near the Fermi energy, which would enhance transport properties. Therefore, the question about how strong correlations affect nonlinear responses is rather nontrivial. A recent general analysis of strong correlation effects on response functions in a simple model has shown that strong correlations should enhance the nonlinear response\cite{michishita_2020}.
However, a demonstration of this enhancement on a concrete model of a strongly correlated system is yet missing.

Thus, in this paper, we study the Weyl-Kondo semimetal $\cebipd$, and we examine the relationship between strong correlations and nonlinear responses.
We adopt a periodic Anderson model which corresponds to a Weyl-Kondo semimetal and analyze it using dynamical mean-field theory(DMFT)\cite{Georges_1996},  which takes local correlation effects and the Kondo effect into account. We calculate the nonlinear longitudinal conductivity and the nonlinear Hall conductivity, and we clarify their temperature dependence.

Our analysis reveals that the nonlinear Hall effect is strongly enhanced in Weyl-Kondo semimetals due to the renormalization of the band structure. 

The rest of this paper is organized as follows: In section II, we introduce the model and methods adopted in our study.
In section III, we show our numerical results. First, we study the spectral functions and the emergence of Weyl-points when decreasing the temperature.
Then, utilizing these results, we calculate the linear and nonlinear conductivities and clarify their temperature dependence.
In section IV, we compare our results with the experimental observations in $\cebipd$.
Finally, we summarize our results.

\section{Model \& Methods}

\subsection{Periodic Anderson Model}
\label{pam}

\begin{figure}[t]
\begin{center}
\includegraphics[width=0.99\linewidth]{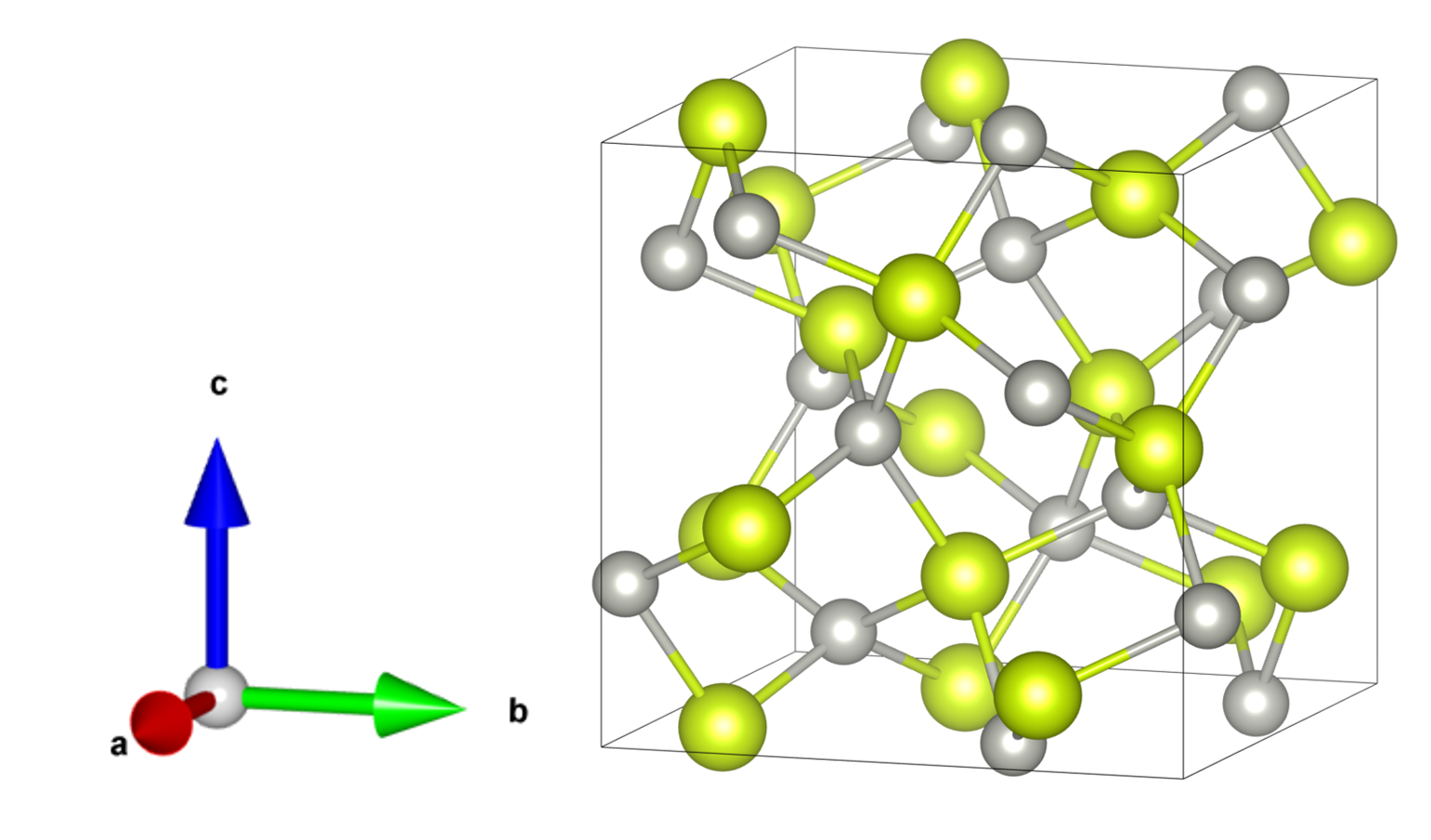}
\end{center}
    \caption{Crystal structure of $\mathrm{Ce_{3} Bi_{4} Pd_{3}}$ visualized with VESTA\cite{Momma_2011} focusing on the $\mathrm{Ce}$(yellow) and $\mathrm{Pd}$(gray) atoms. Note that the Bi atoms are omitted for simplicity.}
    \label{lattice}
\end{figure}
Our primary purpose is to study the low-energy transport properties of a Weyl-Kondo semimetal. 
Transport properties at low temperatures are dominated by the energetically low-lying conduction ($c$) electrons, the localized $f$ electrons, and the Weyl nodes caused by the hybridization and the Kondo effect. 
Thus, motivated by the observation of a giant nonlinear Hall effect in $\cebipd$, we construct a  model, consisting only of one spinful conduction electron band and one spinful $f$ electron band, with the same symmetry as $\cebipd$ hosting Weyl nodes. The conduction electrons and the $f$ electrons in our model correspond to the $4d$ electrons of the Pd atoms and the $4f$ electrons of the Ce atoms in $\cebipd$ respectively. The crystal structure of $\cebipd$ is shown in Fig.~\ref{lattice}. It is a noncentrosymmetric material and is in the nonsymmorphic space group symmetry $T_{d}^{6}$.
The absence of inversion symmetry results in the emergence of the Dresselhaus spin-orbit coupling.
The details of the Hamiltonian are as follows:
\begin{eqnarray}
\label{anderson}
\hat{H}&=&\sum_{\bm{k}} \psi^{\dagger}_{\bm{k}}
H_0(\bm{k})
\psi_{\bm{k}} 
+U \sum_{i} n_{f i \uparrow} n_{f i \downarrow}\\
H_{0} (\bm{k})
&=&\begin{pmatrix}
H_{c} (\bm{k}) &
H_{cf} (\bm{k})\\
H_{f c} (\bm{k}) &
H_{f} (\bm{k})
\end{pmatrix}
\quad \psi^T_{\bm{k}} = 
\begin{pmatrix}
c_{\bm{k}\uparrow}
c_{\bm{k}\downarrow}
f_{\bm{k}\uparrow}
f_{\bm{k}\downarrow}
\end{pmatrix},
\end{eqnarray}
where
\begin{equation}
\label{diago}
H_{c,f} (\bm{k}) = \varepsilon_{c,f} (\bm{k}) \sigma_{0}
+
\bm{D}_{c,f} (\bm{k}) \cdot \bm{\sigma}
\end{equation}
and
\begin{equation}
\label{non_diago}
H_{cf} (\bm{k}) = V({\bm{k}}) \sigma_{0}.
\end{equation}
Here, $c^{\dagger} _{\bm{k}\sigma}(f^{\dagger} _{\bm{k}\sigma})$ creates a $c$ ($f$) electron with momentum $\bm{k}$ and spin $\sigma$. $n_{f i \sigma}$ is the occupation number operator for an $f$ electron with spin $\sigma$ at site $i$. 
The first term in Eq. \eqref{anderson} corresponds to the intra-orbital hopping of the $c$ and $f$ electrons ($H_c$ and $H_f$) and the hybridization between $c$ electrons and $f$ electrons ($H_{cf}$). The second term corresponds to the local Coulomb repulsion between $f$ electrons, where $U>0$ is the strength of the Coulomb repulsion.
The diagonal part of the $c$ and $f$ electron hopping, Eq. \eqref{diago},  consists of two parts.
The first term represents the nearest-neighbor hopping of the electrons, the second term represents the Dresselhaus spin-orbit coupling allowed by the point group symmetry of $\mathrm{Ce_{3} Bi_{4} Pd_{3}}$. 
The specific expressions are given as
\begin{widetext}
\begin{align}
\varepsilon_{c,f} (\bm{k}) &= 2t_{c,f} \{\cos (\bm{k} \cdot \bm{a}_{1}) + \cos (\bm{k} \cdot \bm{a}_{2}) +
\cos (\bm{k} \cdot \bm{a}_{3})\}
+\mu_{c,f}
\\
(\bm{D}_{c,f} (\bm{k}))_{x} &= 
2\lambda_{c,f} \{\sin (\bm{k} \cdot \bm{a}_{3}) - \sin (\bm{k} \cdot \bm{a}_{1})
+
\sin (\bm{k} \cdot (\bm{a}_{3} - \bm{a}_{2}))
+
\sin (\bm{k} \cdot (\bm{a}_{1} - \bm{a}_{2}))\}
\\
(\bm{D}_{c,f} (\bm{k}))_{y} &= 
(\bm{D}_{c,f} (\bm{k};\bm{a}_{1} \rightarrow \bm{a}_{2},\bm{a}_{2} \rightarrow \bm{a}_{3},\bm{a}_{3} \rightarrow \bm{a}_{1}))_{x}
\\
(\bm{D}_{c,f} (\bm{k}))_{z} &= 
(\bm{D}_{c,f} (\bm{k};\bm{a}_{1} \rightarrow \bm{a}_{3},\bm{a}_{2} \rightarrow \bm{a}_{1},\bm{a}_{3} \rightarrow \bm{a}_{2}))_{x}.
\end{align}
\end{widetext}

Here, $\bm{a}_{1} = (a,a,0)$, $\bm{a}_{2} = (0,a,a)$, $\bm{a}_{3} = (a,0,a)$ are the primitive lattice vectors for the face-centered cubic lattice. In this paper, we adopt natural units, where $\hbar=c=k_{\mathrm{B}}=|e|=1$. Furthermore, we set the lattice constant of $\cebipd$ as $a=1$.

The hybridization between the $c$ electrons and $f$ electrons, Eq. $\eqref{non_diago}$, is given by the nearest-neighbor hybridization between $c$ electrons and $f$ electrons, and reads 
\begin{equation}
V (\bm{k}) = 
V (1+e^{i \bm{k} \cdot \bm{a}_{1}} + e^{i \bm{k} \cdot \bm{a}_{2}} +
e^{i \bm{k} \cdot \bm{a}_{3}})  e^{-i \bm{k} \cdot \bm{a}_{4}}.
\end{equation}
Here, we define $\bm{a}_{4}=\frac{1}{2} (\bm{a}_{1}+\bm{a}_{2}+\bm{a}_{3})$ which connects the site of a $c$ electron with that of an $f$ electron in the primitive cell.

\begin{figure}[t]
\begin{center}
\includegraphics[width=0.49\linewidth]{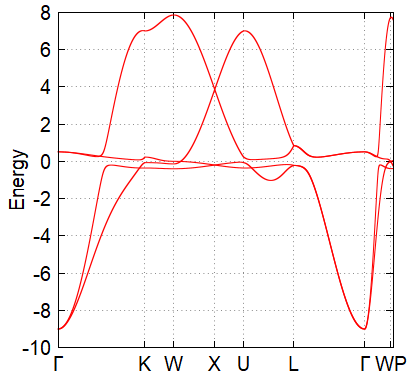}
\includegraphics[width=0.49\linewidth]{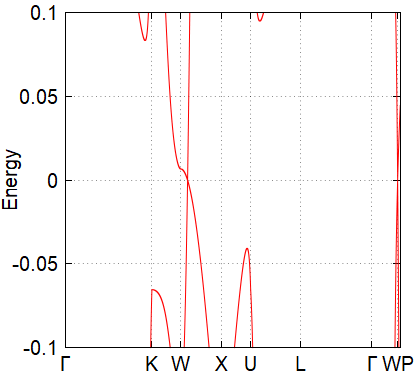}
\end{center}
\caption{Energy dispersion of the free Hamiltonian along the high-symmetry lines in the Brillouin zone (left) and the magnification around  the Fermi energy (right). 
The parameters are $t_{c} = -1.6$, $t_{f} = 0.08$, $\mu_{c}=0.7$, $\mu_{f}=0$, $\lambda_{c}=-1.0$, $\lambda_{f}=0.05$, $V=-0.2$.
We use the following notation for the momenta $\Gamma:(0,0,0,)$, $K:(3 \pi /4,0,3 \pi /4)$, $W:(\pi ,0,\pi /2)$, $X:(\pi,0,0 )$, $U:(\pi,\pi /4,\pi /4 )$, $L:(\pi /2,\pi /2,\pi /2 )$)}
\label{dispersion_1}
\end{figure}

\begin{figure}[t]
\begin{center}
\includegraphics[width=0.49\linewidth]{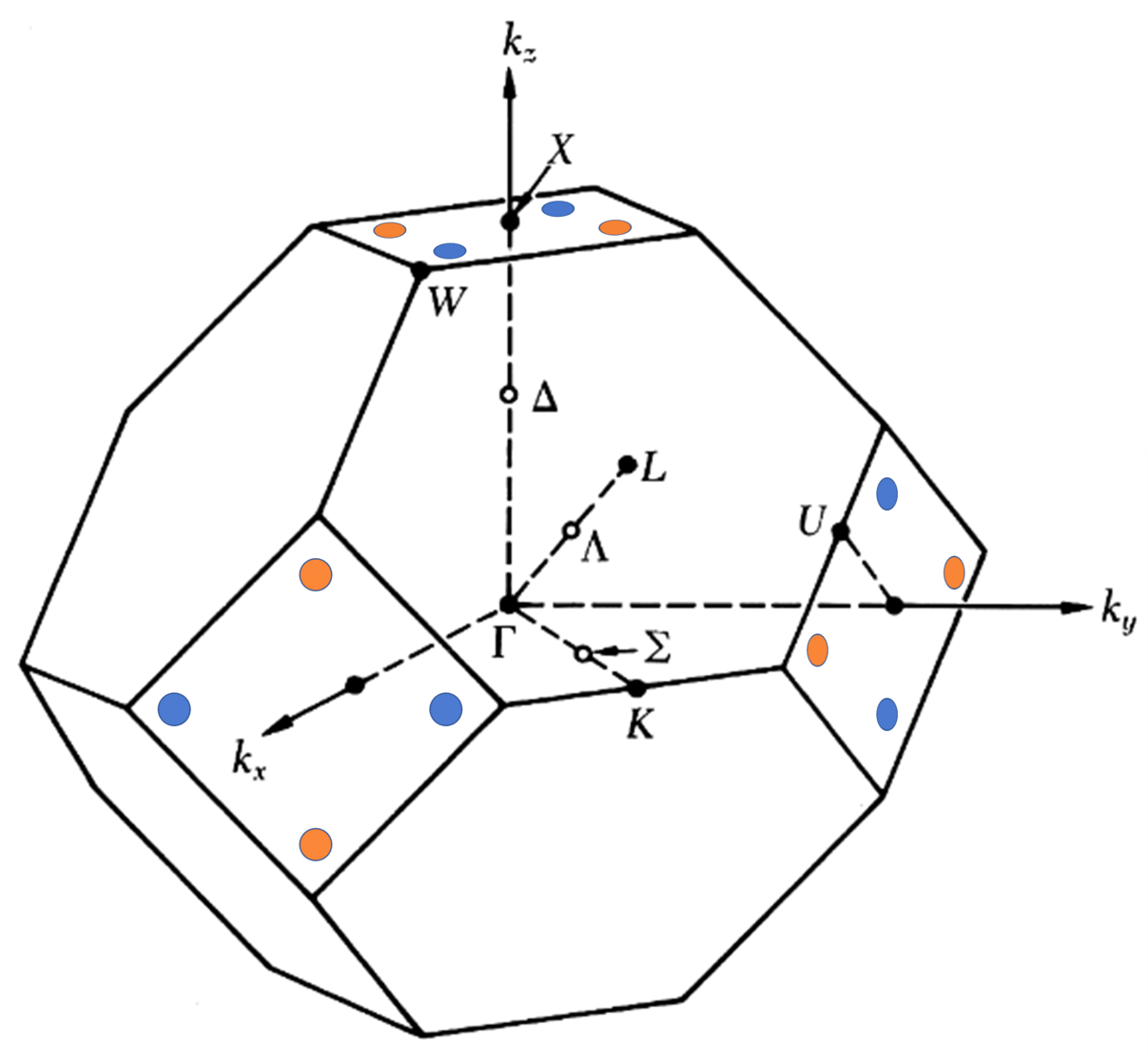}
\includegraphics[width=0.49\linewidth]{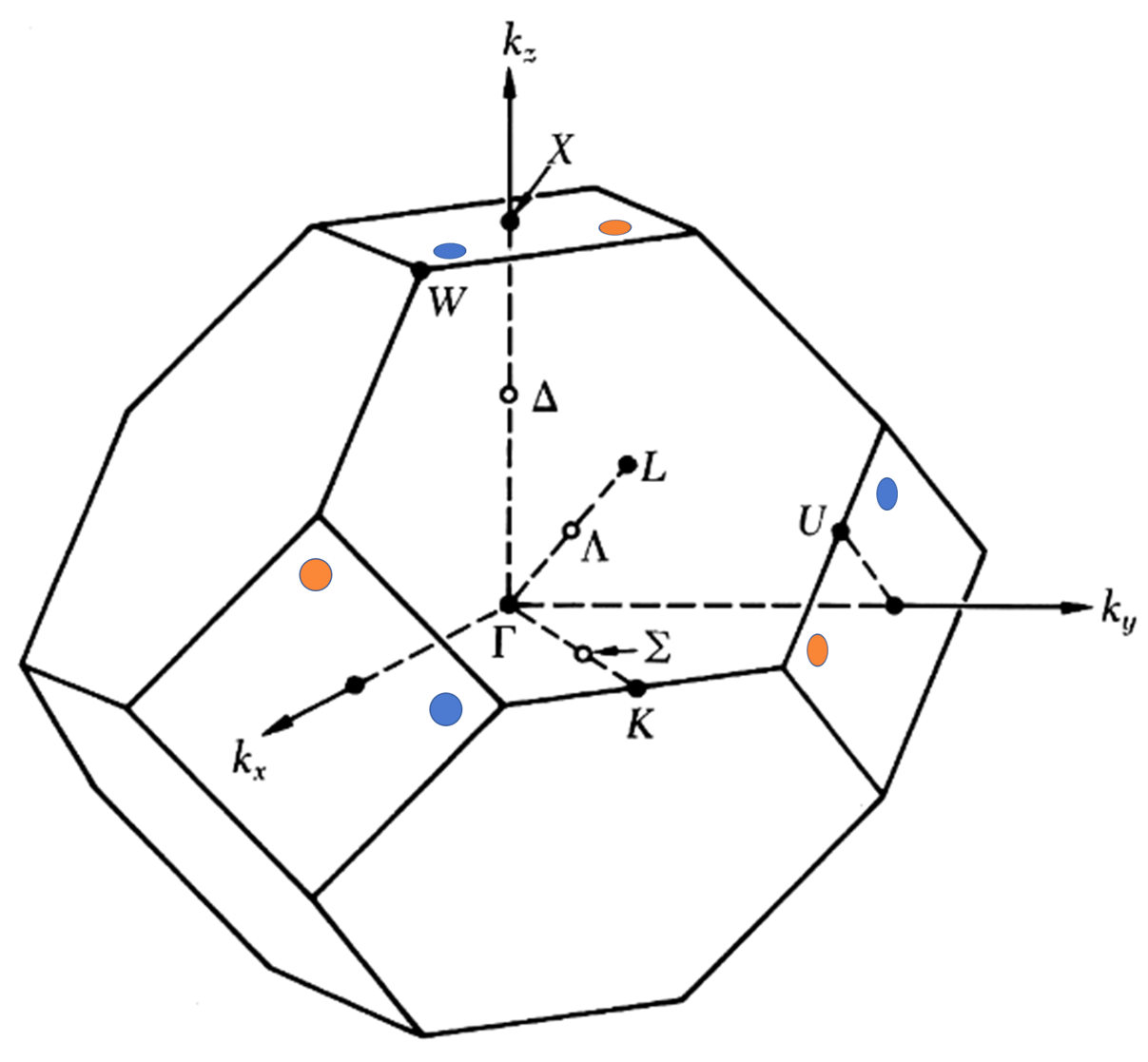}
\end{center}
\caption{Positions of the Weyl-points: The position of the 
Weyl points in the original Hamiltonian point group symmetry ($T_{d}$) (left) and of the Hamiltonian with additional Rashba spin-orbit coupling (right) is shown.
\label{weylpoints_1}}
\end{figure}

In Fig.~\ref{dispersion_1}, we show the energy dispersion of the noninteracting Hamiltonian for the following parameter: $t_{c} = -1.6$, $t_{f} = 0.08$, $\mu_{c}=0.7$, $\lambda_{c}=-1.0$, $\lambda_{f}=-0.05$, $V=-0.2$. 
The eigenvalues of the Hamiltonian are plotted between the high-symmetry points of the fcc lattice and the Weyl point, denoted by WP.
There are $12$ Weyl points near the Fermi energy induced by the hybridization between the $c$ electrons and the $f$ electrons in this model.
The Weyl points lie on the high-symmetry line connecting the X point and the W point in the Brillouin zone, which is consistent with the model constructed by H. Lai $\it{et}$ $\it{al.}$\cite{qimiao_si_2018}. The positions of the Weyl points in the BZ are schematically shown in Fig.~\ref{weylpoints_1} (left panel).

The symmetry of the system poses strict restrictions to the nonlinear Hall effect, as has been generally shown by S. Nandy et al.\cite{nandy_2019}.
While the above-described model corresponds to a Weyl-Kondo semimetal with point group symmetry of $\mathrm{Ce_{3}Bi_{4}Pd_{3}}$ ($T_{d}$), we note that this point group symmetry cannot yield a finite nonlinear Hall conductivity.
In this symmetry group, contributions of the Weyl points to the nonlinear Hall conductivity will vanish.  The symmetry of the Hamiltonian must be further lowered to observe a finite nonlinear Hall resistivity. Because a nonlinear Hall resistivity has been observed in $\cebipd$, we here lower the symmetry of this model by including a Rashba spin-orbit interaction.
 
We add the following Rashba spin-orbit coupling term to the Hamiltonian $H_{c,f} (\bm{k}) \rightarrow H_{c,f} (\bm{k})+ \bm{D}^{R}_{c,f} (\bm{k})\cdot \bm{\sigma}$

\begin{equation}
{\bm{D}^{R}_{c,f}} (\bm{k}) = \bm{v}_{c}(\bm{k}) \times \bm{n}^{R},
\end{equation}
where $\bm{n}^{R}$ is the unit-vector describing the reduction of the symmetry.
 $\bm{v}_{c} (\bm{k})$ is given as
\begin{widetext}
\begin{equation}
(\bm{v}_{c} (\bm{k}))_{x} =
\lambda^{R}_{c,f} \{- \sin (\bm{k} \cdot \bm{a}_{1}) - \sin (\bm{k} \cdot \bm{a}_{3}) -
\sin (\bm{k} \cdot (\bm{a}_{1} - \bm{a}_{2}))
+
\sin (\bm{k} \cdot (\bm{a}_{2} - \bm{a}_{3}))\}
\end{equation}
\begin{equation}
(\bm{v}_{c} (\bm{k}))_{y} = (\bm{v}_{c} (\bm{k};\bm{a}_{1} \rightarrow \bm{a}_{2},\bm{a}_{2} \rightarrow \bm{a}_{3},\bm{a}_{3} \rightarrow \bm{a}_{1}))_{x} 
\end{equation}
\begin{equation}
(\bm{v}_{c} (\bm{k}))_{z} = (\bm{v}_{c} (\bm{k};\bm{a}_{1} \rightarrow \bm{a}_{3},\bm{a}_{2} \rightarrow \bm{a}_{1},\bm{a}_{3} \rightarrow \bm{a}_{2}))_{x}.
\end{equation}
\end{widetext}
We also add the Rashba spin-orbit coupling to the hybridization, $H_{cf} (\bm{k}) \rightarrow H_{cf} (\bm{k})+\bm{V}^{R}_{cf} (\bm{k})\cdot \bm{\sigma} $, which becomes
\begin{equation}
{\bm{V}^{R}_{cf}} (\bm{k}) = \bm{v}_{cf}(\bm{k}) \times \bm{n}^{R},
\end{equation}
where $\bm{v}_{cf} (\bm{k})$ is given as
\begin{equation}
(\bm{v}_{cf} (\bm{k}))_{x} = \frac{i \lambda^{R}_{cf}}{2} (-1+e^{i \bm{k} \cdot \bm{a}_{1}} - e^{i \bm{k} \cdot \bm{a}_{2}} +
e^{i \bm{k} \cdot \bm{a}_{3}}) e^{-i \bm{k} \cdot \bm{a}_{4}}
\end{equation}
\begin{equation}
(\bm{v}_{cf} (\bm{k}))_{y} = (\bm{v}_{cf}(\bm{k};\bm{a}_{1} \rightarrow \bm{a}_{2},\bm{a}_{2} \rightarrow \bm{a}_{3},\bm{a}_{3} \rightarrow \bm{a}_{1}))_{x} 
\end{equation}
\begin{equation}
(\bm{v}_{cf} (\bm{k}))_{z} = (\bm{v}_{cf} (\bm{k};\bm{a}_{1} \rightarrow \bm{a}_{3},\bm{a}_{2} \rightarrow \bm{a}_{1},\bm{a}_{3} \rightarrow \bm{a}_{2}))_{x}.
\end{equation}
All terms in the Hamiltonian which include the superscript "R" originate from an additional reduction of the symmetry and depend on the direction $\bm{n}^{R}$.
In this paper, we will consider an additional symmetry reduction using $\bm{n}^{R}\parallel[100]$ and $\bm{n}^{R}\parallel[111]$.

When including the Rashba spin-orbit coupling to our model, $6$ of the $12$ Weyl points gap out because of the reduction of the symmetry.
In Fig.~\ref{weylpoints_1}, we schematically compare the positions of the Weyl points without further symetry breaking (left panel) with the model including the Rashba spin-orbit coupling (right panel). We note that the positions of the Weyl points for directions $\bm{n}^{R}\parallel[100]$ and $\bm{n}^{R}\parallel[111]$ are nearly the same and cannot be distinguished in this figure.
Furthermore, we note that the Weyl points do
no longer appear on the boundary of the BZ when the Rashba terms are included.

As already stated above, the symmetry of the system poses strict restrictions to the nonlinear Hall effect.
We can now analyze the symmetry of the system, including the additional Rashba terms.
When $\bm{n}^{R}\parallel[111]$, the symmetry of the system reduces to $C_{3v}$, which does not allow for a nonlinear Hall effect.
On the other hand, when $\bm{n}^{R}\parallel[100]$, the symmetry of the system reduces to $C_{1h}$, which is low enough to realize a nonlinear Hall effect.
To study the nonlinear Hall effect, we do not have to consider the case, $\bm{n}^{R} \parallel [111]$. However, to show that the symmetry reduction does not have a crucial influence on other responses except for the nonlinear Hall effect, we also consider $\bm{n}^{R} \parallel [111]$ in this paper.

While we do not want to specify the cause of the additional Rashba spin-orbit coupling, the experimental observation of the nonlinear Hall effect confirms the symmetry reduction.
In the experiment\cite{Dzsaber_2021}, the observation of the Hall effect has been explained by the reduction of the symmetry through an electric field. Thus, the external electric field leads to a non-perturbative effect, which must be included in the Hamiltonian. Such a non-perturbative effect might be feasible because the Fermi surface of a Weyl-semimetal is tiny. However, because other mechanisms might exist, we here include the symmetry reduction as a general Rashba spin-orbit interaction with direction dependence $\bm{n}^{R}$.

Finally, we include correlation effects in this model by using the dynamical mean-field theory (DMFT)\cite{Georges_1996}. 
DMFT takes local correlation effects fully into account by mapping the lattice model onto a quantum impurity model. To calculate Green's functions and self-energies for the resulting quantum impurity model, we use in this paper the numerical renormalization group (NRG)\cite{NRG_review}.
NRG is an accurate and reliable numerical method to study the low-energy properties of impurity models. A big advantage of NRG is the ability to calculate self-energies for real-frequencies without analytical continuation.
Thus, DMFT+NRG can be expected to be a suitable approach to elucidate strong correlation effects on low-energy transport phenomena.

\subsection{Nonlinear Conductivity}
After having calculated a self-consistent self-energy and Green's functions for the interacting model using DMFT, we will calculate linear and nonlinear conductivities using the Kubo formalism.

\begin{figure*}[t]
\begin{center}
\includegraphics[width=0.32\linewidth]{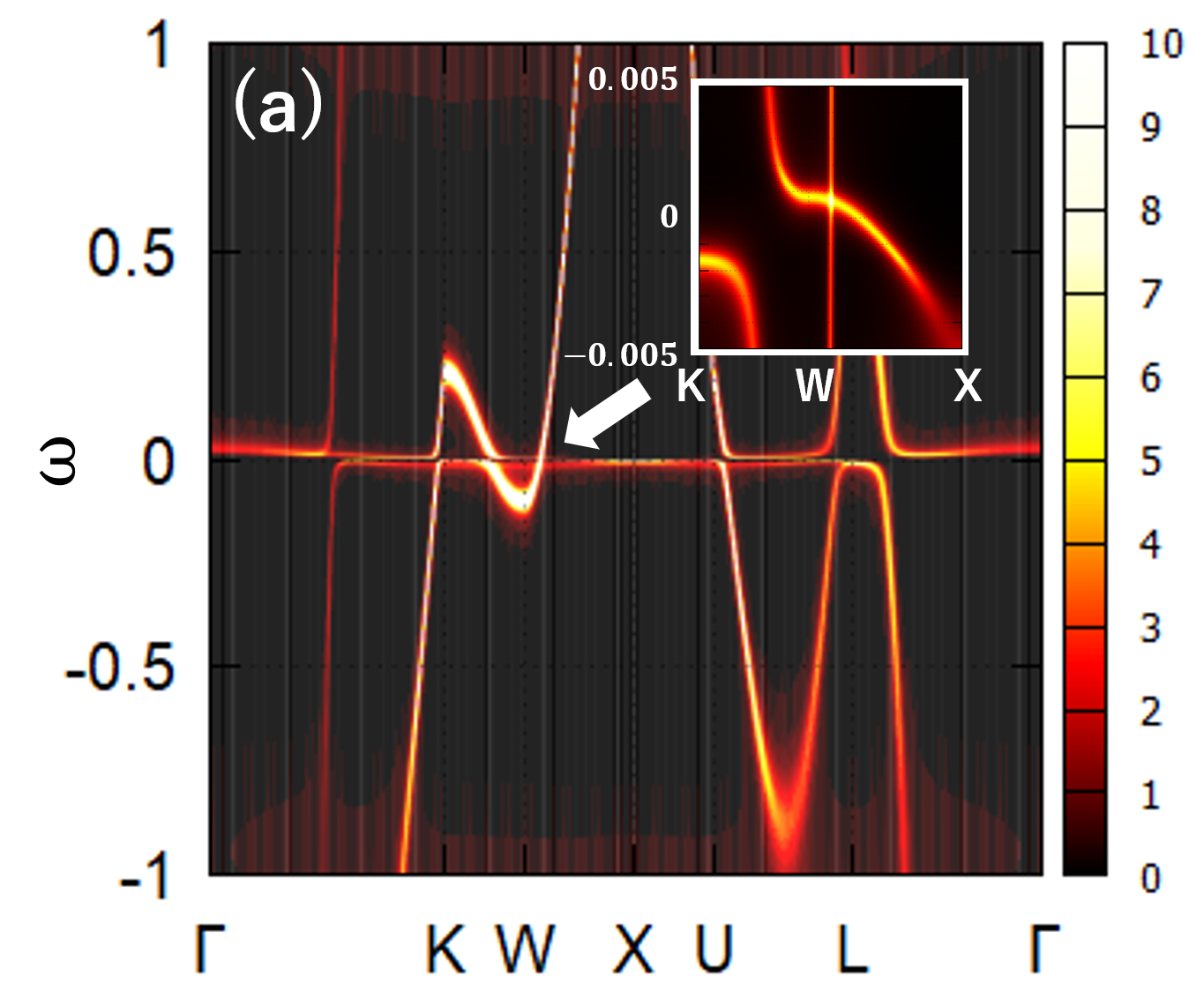}
\includegraphics[width=0.32\linewidth]{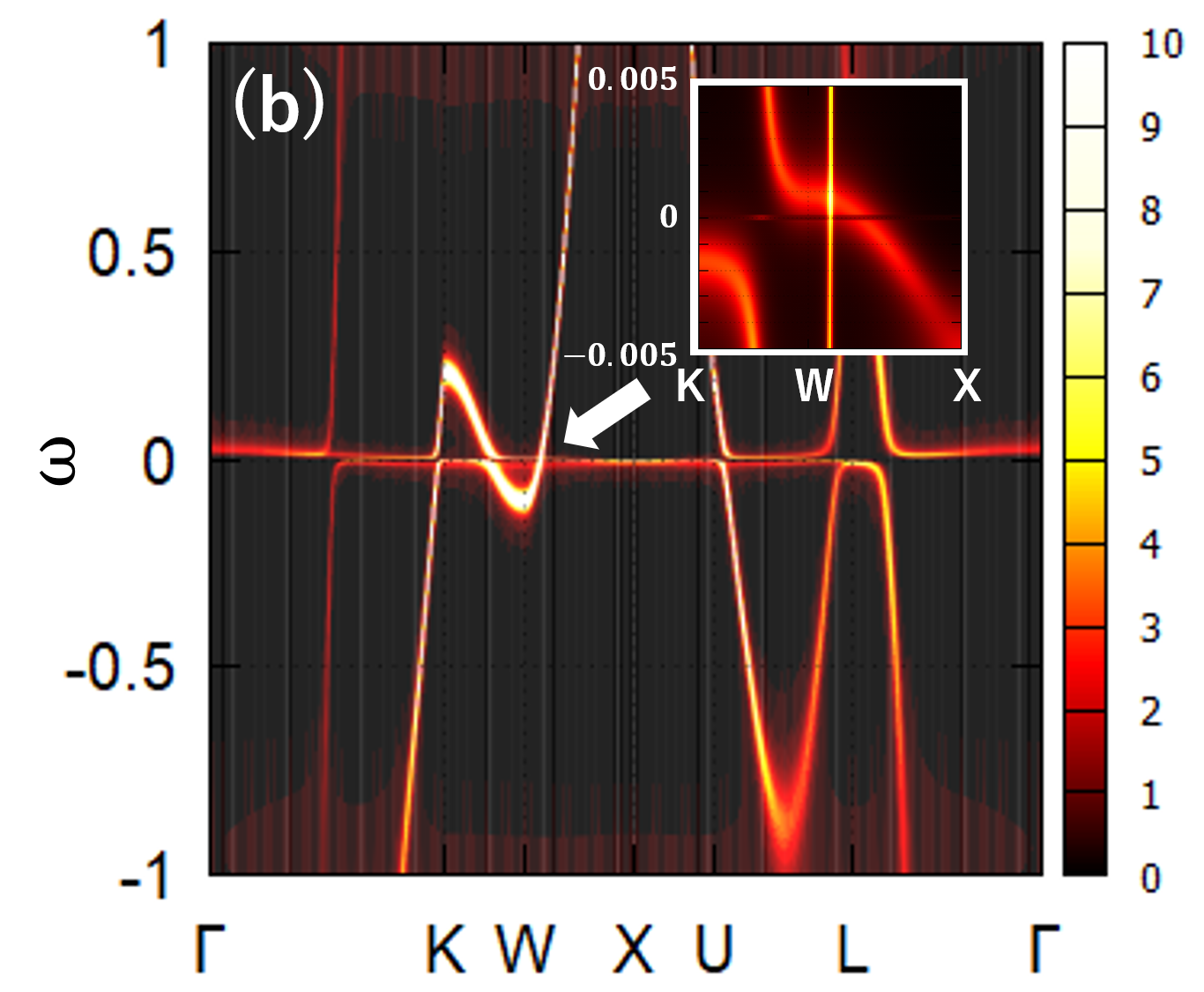}
\includegraphics[width=0.32\linewidth]{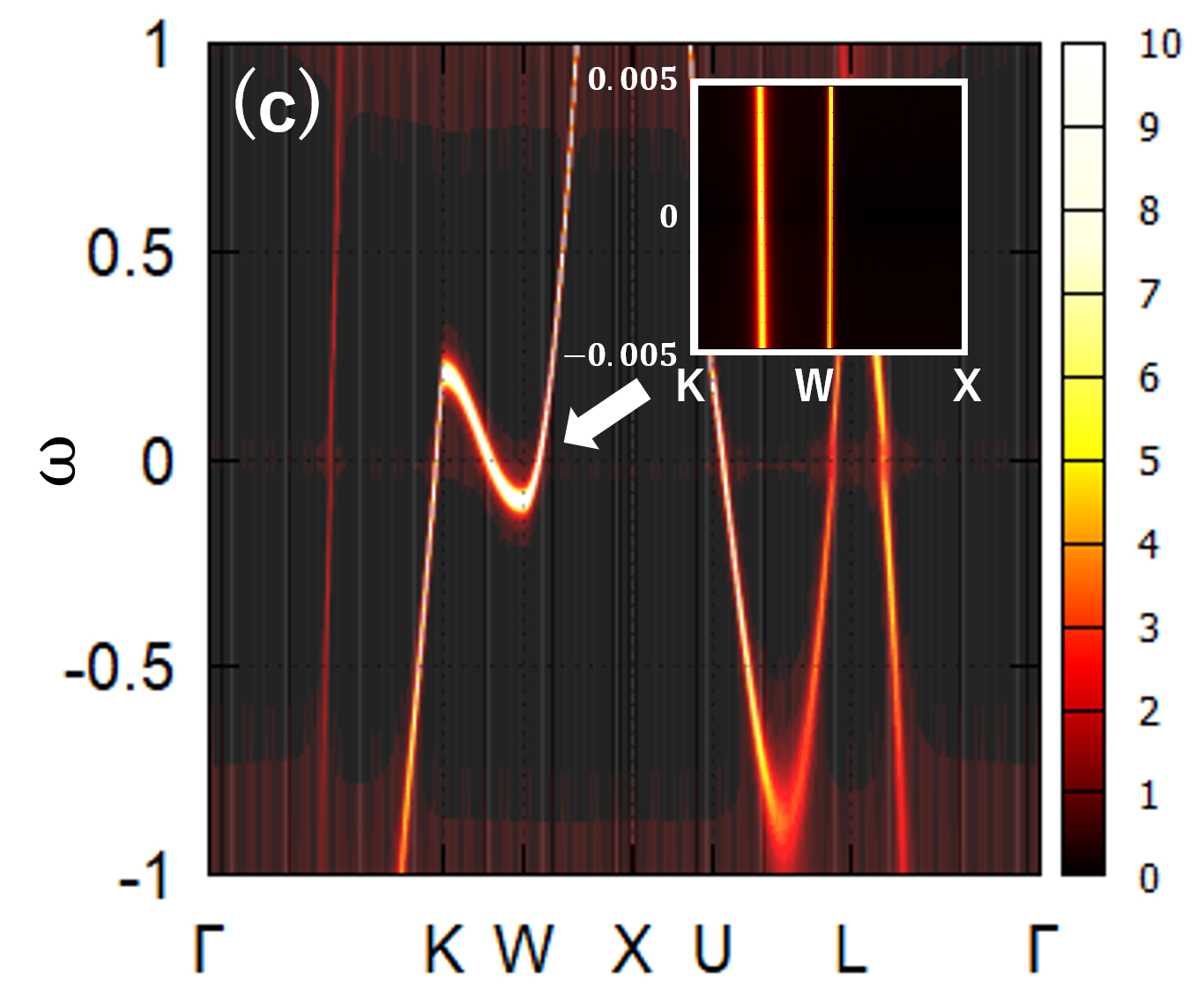}
\end{center}
\caption{Spectral function for different temperatures (from left to right, $T=0.0001$, $T=0.0016$ and $T=0.0256$). The insets show magnifications around the Weyl-points.}
\label{spectrum}
\end{figure*}

Using the velocity gauge, we can derive a formula for the nonlinear conductivity, which is only based on single-particle Green's functions neglecting vertex corrections\cite{parker_2019,lopes_2019,michishita_2020}. According to the work by Y. Michishita and R. Peters\cite{michishita_2020}, the nonlinear conductivity reads
\begin{widetext}
\begin{equation}
\label{nonlinear_Kubo}
\begin{split}
\sigma_{i j k} &= -2\int_{B Z} \frac{d^{3} k}{4 \pi^{3}} \int \frac{d \omega}{2 \pi} (-\frac{\del f( \omega)}{\del \omega}) \mathrm{Im} (\mathrm{Tr} [J_{i}(\bm{k}) \frac{\del G^{R}(\bm{k}, \omega)}{\del \omega} J_{j}(\bm{k}) G^{R}(\bm{k}, \omega) J_{k}(\bm{k}) G^{A}(\bm{k}, \omega)]) 
\\
&-\int_{B Z} \frac{d^{3} k}{4 \pi^{3}} \int \frac{d \omega}{2 \pi} (-\frac{\del f( \omega)}{\del \omega}) \mathrm{Im} (\mathrm{Tr} [J_{i}(\bm{k}) \frac{\del G^{R}(\bm{k}, \omega)}{\del \omega} J_{j k}(\bm{k}) G^{A}(\bm{k}, \omega)] )
\\
&+ 2\int_{B Z} \frac{d^{3} k}{4 \pi^{3}} \int \frac{d \omega}{2 \pi} f (\omega) \mathrm{Im} (\mathrm{Tr} [J_{i}(\bm{k}) \frac{\del}{\del \omega} (\frac{\del G^{R}(\bm{k}, \omega)}{\del \omega} J_{j}(\bm{k}) G^{R}(\bm{k}, \omega)) J_{k}(\bm{k}) G^{R}(\bm{k}, \omega)])
\\
&+ \int_{B Z} \frac{d^{3} k}{4 \pi^{3}} \int \frac{d \omega}{2 \pi} f (\omega) \mathrm{Im} (\mathrm{Tr} [J_i(\bm{k}) \frac{{\del}^{2} G^{R}(\bm{k}, \omega)}{\del {\omega}^{2}} J_{j k}(\bm{k}) G^{R}(\bm{k}, \omega)])+(j\leftrightarrow k)
\end{split}
\end{equation}
\end{widetext}
$\sigma_{ijk}$ is the nonlinear conductivity when the external electric field is applied in direction $j$ and $k$ and the current is induced along direction $i$.
Here, $f (\omega)$ is the Fermi distribution function, $G_{R} (G_{A})$ is the retarded(advanced) Green's function of the system and $\mathrm{Tr}$ indicates the trace over the orbital degrees of freedom ($c_\uparrow ,c_\downarrow,f_\uparrow,f_\downarrow$ in this model).
$J_{i}$ and $J_{ij}$ are velocities, defined as 
\[
J_{i}(\bm{k}) = -\frac{\del H(\bm{k})}{\del k_{i}} \qquad J_{i j}(\bm{k}) = \frac{{\del}^{2} H(\bm{k})}{\del k_{i}\del k_{j}},
\]
where the minus sign arises due to the negative charge of the electron.
In particular, as for the nonlinear Hall conductivity, which has been shown to originate from the Berry curvature dipole of the system by Inti Sodemann and Liang Fu\cite{Sodemann_2015} in a semi-classical treatment, we note that the first term of Eq. \eqref{nonlinear_Kubo} contains the contribution from the Berry curvature dipole.
We note that the above expression \eqref{nonlinear_Kubo} is derived in a fully quantum mechanical treatment. Thus, this formula includes other terms representing inter-band transitions and contributions from the Fermi sea, which are absent in a semi-classical treatment using the Boltzmann equation.

\section{Results}
\label{results}

\subsection{Without further symmetry reduction}
\label{E_is_not_considered}
\subsubsection{Spectral functions}
\begin{figure}[t]
\begin{center}
\includegraphics[width=0.49\linewidth]{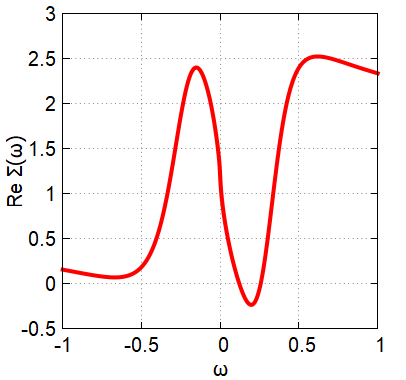}
\includegraphics[width=0.49\linewidth]{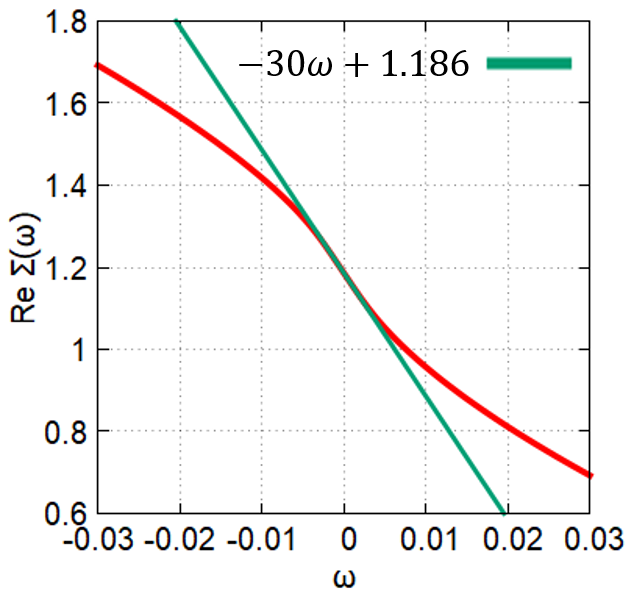}
\end{center}
\caption{Real part of the self-energy at $T=0.0001$ in the system without further symmetry reduction. The right panel shows a magnification around the Fermi energy including a linear fit.}
\label{selfenergy_000}
\end{figure}

In this section, we first analyze correlation effects in the Weyl-Kondo semimetal without further symmetry reduction, i.e. without including the Rashba spin-orbit coupling. We use the following parameters: $t_{c} = -1.6$, $t_{f}=0.08$, $\mu_{c}=0.7$, $\mu_{f}=-1.2$, $\lambda_{c}=-1.0$, $\lambda_{f}=-0.05$, $V=-0.2$, $U=2.4$, $\lambda_{c,f}^R=0$, $\lambda_{cf}^R=0$.

In Fig.~\ref{spectrum}, we plot the spectral functions along the high-symmetry lines for three characteristic temperatures, $T=0.0001$, $T=0.0016$, and $T=0.0256$ (from the left to the right).
The spectral function shown in the left panel of Fig.~\ref{spectrum}, calculated for $T=0.0001$, demonstrates the existence of Weyl points in the interacting model close to the Fermi energy. At this temperature, the system is in a Fermi-liquid state so that the imaginary part of the self-energy vanishes quadratically around the Fermi energy. The corresponding real part of the self-energy is shown in Fig.~\ref{selfenergy_000}. As displayed in the right panel of Fig.~\ref{selfenergy_000}, the slope of the real part of the self-energy at $\omega=0$ is about $30$. Thus, the mass of an electron in the vicinity of the Fermi surface is enhanced about $31$ times resulting in a strongly renormalized band structure. 

Increasing the temperature, the $f$ electrons begin to localize, which can be observed in the middle panel of Fig.~\ref{spectrum} at approximately $T=0.0016$. 
Above $T \sim 0.0016$ intense electron-electron scattering near the Fermi surface weakens the hybridization between the $c$ electrons and the $f$ electrons. Because the Weyl points emerge due to the hybridization between $c$ and $f$ electrons, the Weyl points disappear in the spectrum, shown in the right panel of Fig.~\ref{spectrum} at approximately $T=0.0256$. 
Based on calculations of the spectral function for different temperatures, we estimate the Kondo temperature in our model to $T_{\mathrm{K}} \sim 0.0016$.

\subsubsection{Transport properties}
\label{linear_000}

Throughout this paper, we assume a time-reversal symmetric system. Thus, a finite linear Hall conductivity is forbidden by the Onsager's reciprocal theorem. Furthermore, in this calculation without further symmetry reduction, i.e. without Rashba spin-orbit coupling, also the nonlinear Hall effect is forbidden by symmetry.

\begin{figure}[t]
\begin{center}
\includegraphics[width=0.49\linewidth]{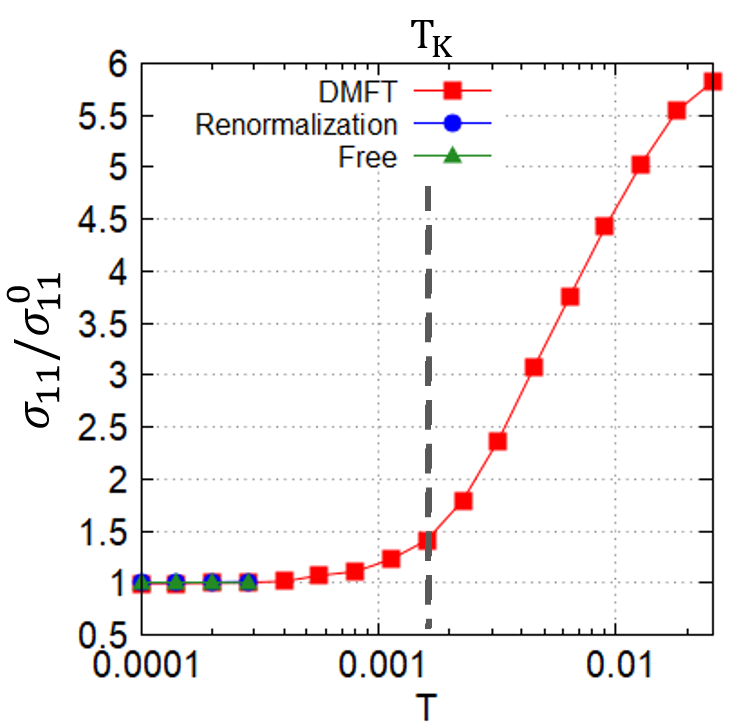}
\includegraphics[width=0.49\linewidth]{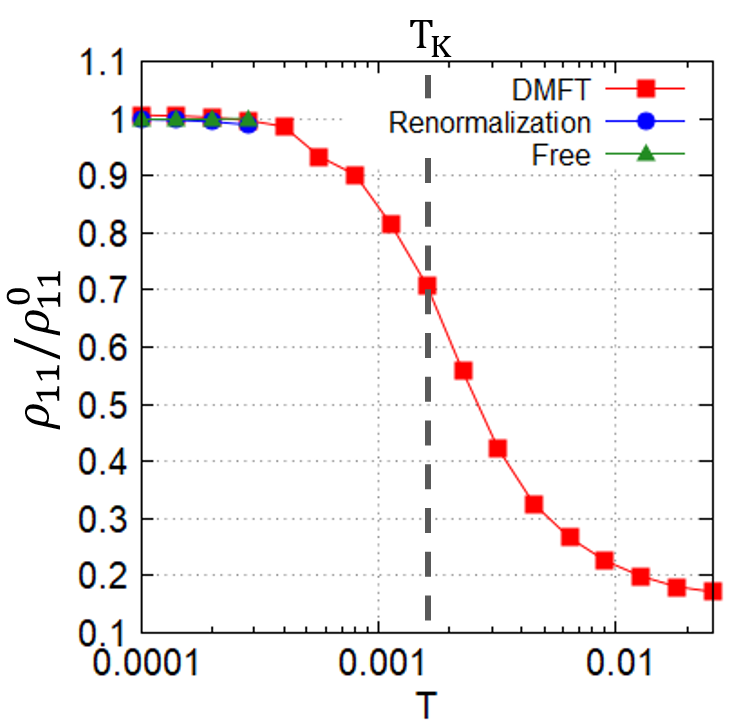}
\end{center}
\caption{Longitudinal linear conductivity $\sigma_{11}$ for the system without further symmetry reduction and the corresponding longitudinal resistivity $\rho_{11}$. The conductivity and the resistivity are normalized by the conductivity and the resistivity at $T=0.0001$ of the noninteracting (Free) system, $\sigma_{11}^{0}=0.573$ and $\rho_{11}^{0}=1.75$ respectively. The vertical dashed line corresponds to the Kondo temperature read off from the spectral functions. The meaning of the different symbols is explained in the main text.}
\label{linear_conductivity}
\end{figure}

In Fig.~\ref{linear_conductivity}, we first show the temperature dependence of the linear longitudinal electrical conductivity and the corresponding resistivity, calculated as the inverse of the linear longitudinal conductivity.
To better understand the effect of strong correlations on the linear longitudinal conductivity, we include in this figure the conductivity calculated with the full self-energy obtained by DMFT, the conductivity of the free system only considering the constant energy shift by the self-energy, $\Sigma(\omega=0,T=0)$, denoted as "Free", and the conductivity calculated considering terms up to linear order $\omega$ in the self-energy around the Fermi energy, which is denoted as "Renormalization." We note that the renormalization of the band structure originates from the linear term of the real part of the self-energy around the Fermi energy. Thus, approximating the self-energy as 
$\Sigma(\omega)=\Sigma(\omega=0,T=0)+\frac{\partial \Sigma(\omega=0,T=0)}{\partial \omega}\omega$
and neglecting the imaginary part which is quadratic in the frequency, the band structure around the Fermi energy appears just as in the full calculation.

As the temperature is lowered below the Kondo temperature $T_{\mathrm{K}}$, the resistivity increases.
At high temperatures, $f$ electrons are localized and absent from the Fermi energy. The system behaves metallic due to the $c$ electrons. 
As has been demonstrated in Fig.~\ref{spectrum}, around the Kondo temperature, $f$ electrons begin to hybridize with the $c$ electrons forming Weyl points close to the Fermi energy. Besides the Weyl points, most of the Fermi surface gaps out at this temperature. 
Generally, the Weyl points in a Weyl-Kondo semimetal lie in the vicinity of the Fermi energy. Thus, the Fermi surface is extremely small. Because the linear longitudinal conductivity is proportional to the density of states at the Fermi surface, the resistivity is large at low temperatures. 

We are now able to compare the linear resistivities obtained by the full self-energy ("DMFT") with that obtained using the linear part of the self-energy ("Renormalization"), and that of the free system ("Free").
Because only the full self-energy can describe the crossover from a metallic state at high temperatures to the Weyl-semimetal at low temperatures, we only compare these resistivities at very low temperatures.
As shown in Fig.~\ref{linear_conductivity}, the magnitude of the linear resistivity at low temperature is nearly the same for all three calculations, "DMFT", "Free", and "Renormalization". Namely, at low temperature in the Fermi liquid state, strong correlations have little effect on the magnitude of the linear resistivity.
This can be understood as follows\cite{michishita_2020}:
Assuming that the self-energies are the same for each orbital, we see that the self-energy is proportional to the identity matrix. Furthermore, assuming that the contribution from the Fermi surface is dominant to the conductivity, one finds the following relation between the conductivities of the interacting and the free system:
\begin{equation}
    \sigma_{ij_{1}\cdots j_{n}}^{int} (\omega_{1},\ldots,\omega_{n}) = \frac{1}{Z^{n-1}} \sigma_{ij_{1}\cdots j_{n}}^{free} (\omega_{1}^{'},\ldots,\omega_{n}^{'}),
\end{equation}
where $Z$ is the renormalization constant for the system, $n$ is the order of the conductivity, and $\omega_{i}^{'}$ is the renormalized input frequency, $\omega_{i}^{'}=\omega_{i}/Z$.
Therefore, strong correlations do not strongly affect the magnitude of the linear conductivity, $n=1$. On the other hand, we already see here that a second-order nonlinear conductivity, $n=2$, can be expected to be enhanced by $1/Z(>1)$ times compared to a noninteracting system.
We have to note that, in the here analyzed periodic Anderson model, where only the f electrons are correlated, the situation is more complicated. However, the results demonstrate that the renormalization does not strongly affect the linear conductivity even in the periodic Anderson model.

\begin{figure}[t]
\begin{center}
\includegraphics[width=0.49\linewidth]{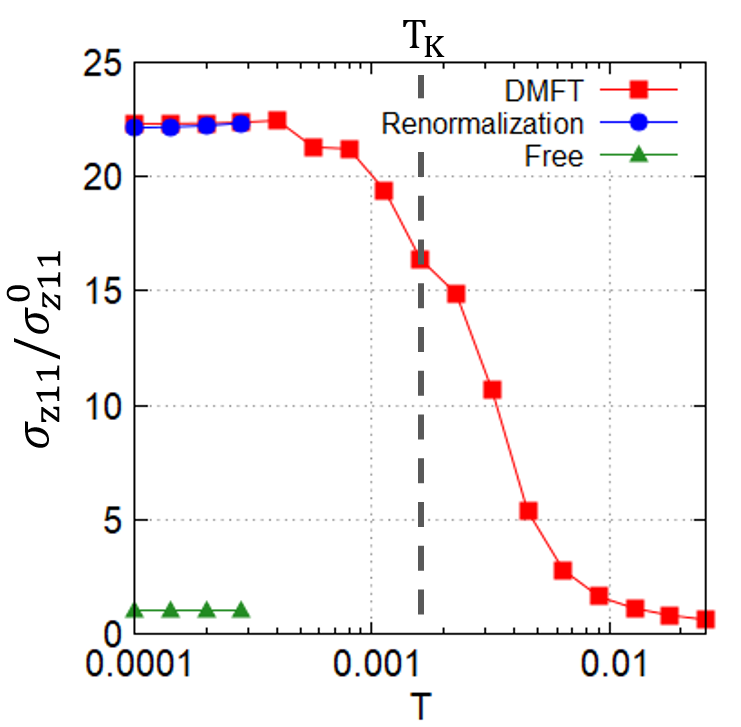}
\includegraphics[width=0.49\linewidth]{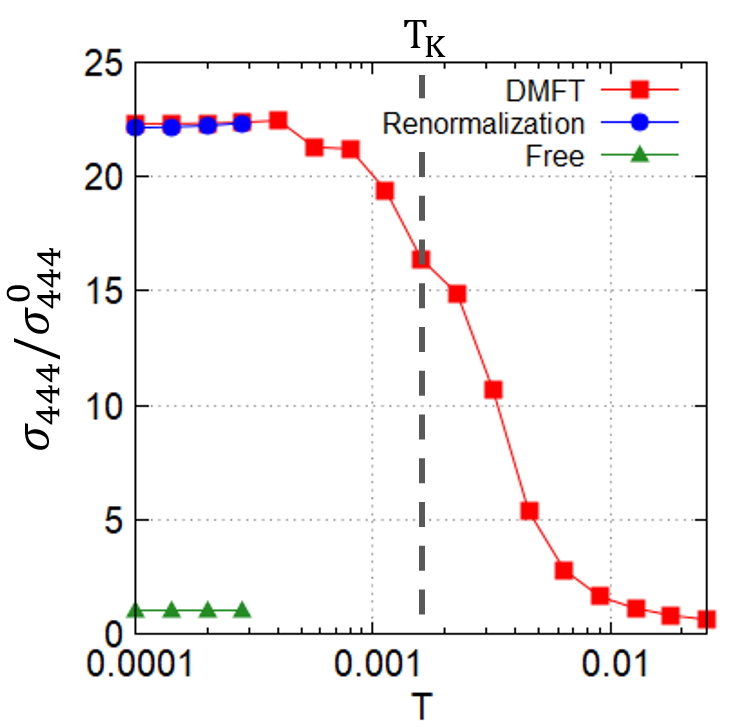}
\end{center}
\caption{Nonlinear transverse conductivity $\sigma_{z11}$ and nonlinear longitudinal conductivity $\sigma_{444}$ without further symmetry reduction. The transverse conductivity and the longitudinal conductivity are normalized by the transverse conductivity  and the longitudinal conductivity at $T=0.0001$ for the noninteracting (Free) system, $\sigma_{z11}^{0}=0.589$ and $\sigma_{444}^{0}=0.677$ respectively.}
\label{nonlinear_conductivity_000}
\end{figure}

Next, let us discuss nonlinear conductivities in the model without Rashba spin-orbit coupling.
In Fig.~\ref{nonlinear_conductivity_000}, we show the temperature dependence of the nonlinear conductivities. 
The left panel of Fig.~\ref{nonlinear_conductivity_000} shows the nonlinear transverse conductivity when the electric field is applied to $[100]$ and the current is measured in the $z$-direction. The right panel shows the nonlinear longitudinal conductivity when the electric field is applied to $[111]$ and the current is also measured in this direction.
We again compare the DMFT results with the "Free" system and the conductivity calculated only including the linear part of the self-energy denoted by "Renormalization".
We note that the transverse conductivity is different from a Hall conductivity, which is defined as the antisymmetric part of the conductivity tensor (see Eq.~\eqref{hall_definition}). A true Hall effect is not allowed in the current system without Rashba spin-orbit coupling because the symmetry of the system $(T_{d})$ is too high. The only finite component of the nonlinear conductivity tensor is $\sigma_{xyz}$, which is easily shown by performing the symmetry operations of $T_{d}$ to an arbitrary nonlinear conductivity $\sigma_{ijk}$.
Thus, both of the nonlinear conductivities shown in Fig.~\ref{nonlinear_conductivity_000}, nonlinear transverse and longitudinal conductivity, have to be proportional to $\sigma_{xyz}$ and thus to each other. Consequently, the temperature dependence of both shown nonlinear conductivites is exactly the same.

As the linear conductivities, 
also the nonlinear conductivities are strongly affected by the Kondo effect.
As can be seen in Fig.~\ref{nonlinear_conductivity_000}, when the temperature is lowered below the Kondo temperature, the nonlinear conductivities increase rapidly.

Next, let us take a look at the renormalization effects on the nonlinear conductivities.
The magnitude of the nonlinear conductivities at low temperatures including the full self-energy ("DMFT") and the linear part of the self-energy ("Renormalization"), are about $22$ times as large as the conductivity of the noninteracting system ("Free").
Because the conductivities denoted by "DMFT" and "Renormalization" have the same magnitude, we understand that the enhancement originates from the renormalized band structure.
Thus, the expectation that nonlinear conductivities are enhanced by renormalization effects holds.
However, although the renormalization enhances the nonlinear conductivity, we already see that the enhancement factor is not 31 as expected from the simple analysis treating interactions in all orbitals equivalent.

With the above considerations in mind, we can build the following physical picture: 
 At high temperatures, $f$ electrons are localized because of the strong Coulomb repulsion. The main contributions to the linear and nonlinear conductivity come from the $c$ electrons close to the Fermi energy.
The enhancement of the nonlinear conductivity due to strong correlation effects does not appear at high temperatures.
On the other hand, as the temperature is lowered, the hybridization between the $c$ electrons and the $f$ electrons becomes essential. The band structure close to the Fermi energy corresponds to that of a strongly renormalized Weyl semimetal. The renormalization of the band structure only affects the nonlinear conductivity, which is enhanced. Therefore, while the linear conductivity decreases due to the formation of a Weyl-semimetal when lowering the temperature, the nonlinear conductivity increases. 

Finally, although it seems that the magnitude of the nonlinear longitudinal conductivity is as large as that of the linear longitudinal conductivity, we note that their dimensions are different. Thus, these conductivities cannot be compared directly. 
To compare them, we have to suppose a magnitude of the electric field and consider the corresponding physical constants, which we have set to 1 in the current paper. For example, suppose that the magnitude of the electric field is $100$ $[\mathrm{V/m}]$. In that case, the nonlinear longitudinal conductivity is $10^{-10}$ smaller than the linear when we consider the necessary physical constants $\frac{|e|a^{2}}{c \hbar}$, which is the order of $10^{-12}$ [m/V].

\subsection{System with further symmetry reduction}
\label{E_is_considered}

We now include a Rashba spin-orbit coupling originating from a further reduction of the symmetry. The parameters describing the Rashba interaction are as follows: $\lambda_{c}^{R} = -0.32,\lambda_{f}^{R} = 0.016$, and  $\lambda_{cf}^{R}=-0.04$.

To check our model's validity, we have calculated the spectral functions, self-energies, and the linear longitudinal resistivity of the model when the symmetry is further reduced.
The symmetry reduction using the above-stated parameters does not strongly affect these properties. As described in the model section, some of the Weyl points gap out.
Furthermore, the renormalization factor, corresponding to the linear term of the self-energy at the Fermi energy, is slightly reduced to ($1/Z \sim 28.5$). 
We show those spectral functions and linear conductivities in appendix \ref{appendix_dmft}. 

\subsubsection{Nonlinear Conductivity}
\label{nonlinear_conductivity_111_100}

\begin{figure}[t]
\begin{center}
\includegraphics[width=0.49\linewidth]{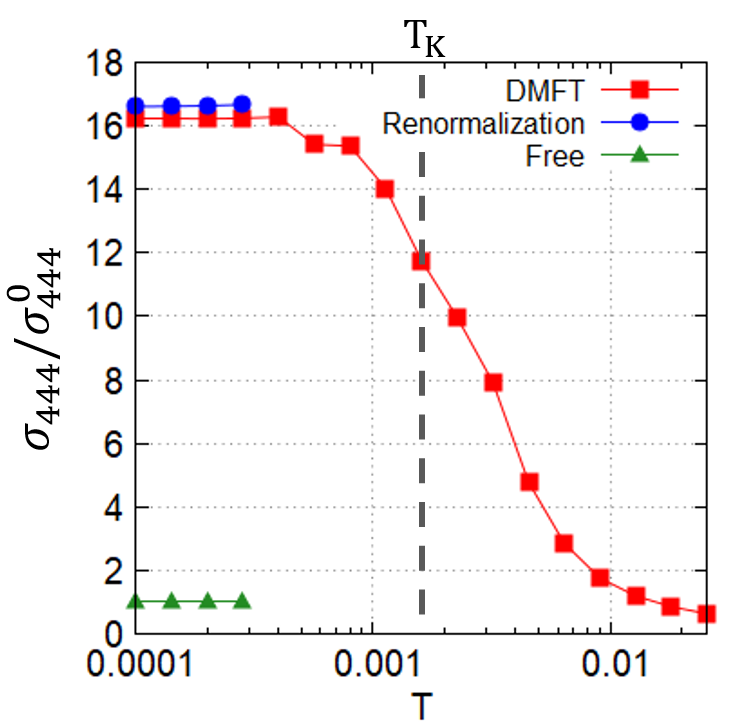}
\includegraphics[width=0.49\linewidth]{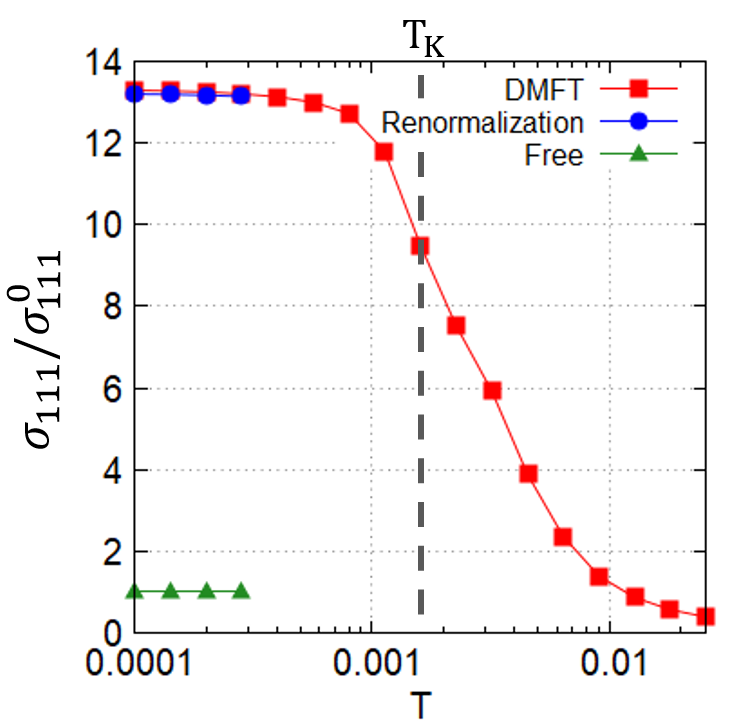}
\end{center}
\caption{Nonlinear longitudinal conductivities for $\bm{n}^{R}  \parallel [111]$ (left) and $\bm{n}^{R} \parallel [100]$ (right). The conductivities are taken in the same direction as the vector of the Rashba spin-orbit coupling. The conductivity for $\bm{n}^{R}  \parallel [111]$ and the conductivity for $\bm{n}^{R} \parallel [100]$ are normalized by the corresponding conductivities at $T=0.0001$ for the noninteracting (Free) system, $\sigma_{444}^{0}=1.07$ and $\sigma_{111}^{0}=0.311$ respectively.} 
\label{nonlinear_conductivity_100}
\end{figure}

Next, let us analyze  nonlinear conductivities.
In Fig.~\ref{nonlinear_conductivity_100}, we show the temperature dependence of the nonlinear longitudinal conductivities.
Without further symmetry reduction, only the nonlinear longitudinal conductivity in the direction $\bm{a}_{4}=\frac{1}{2} (\bm{a}_{1}+\bm{a}_{2}+\bm{a}_{3})$ with an applied electric field in the same direction can be finite.
On the other hand, when the symmetry of the Hamiltonian is further reduced, different nonlinear longitudinal conductivities can be finite.

As can be seen from Fig.~\ref{nonlinear_conductivity_100}, the temperature dependencies of the shown nonlinear longitudinal conductivies are almost the same. The nonlinear conductivity increases rapidly when the temperature is lowered below the Kondo temperature. Comparing between the noninteracting system ("Free") and the calculation including the full self-energy ("DMFT"), we see that the nonlinear conductivity is again strongly enhanced by the correlation effects.

\begin{figure}[t]
\begin{center}
\includegraphics[width=0.49\linewidth]{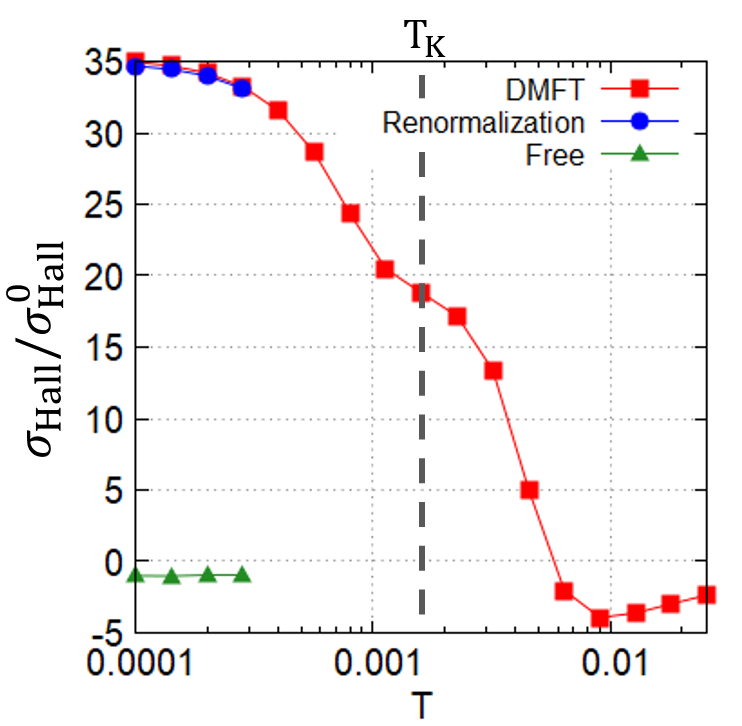}
\includegraphics[width=0.49\linewidth]{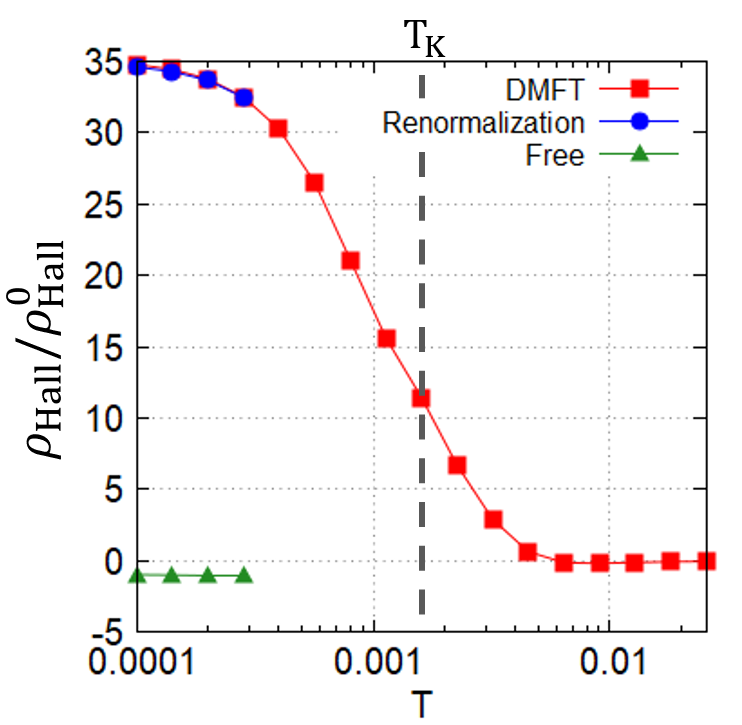}
\end{center}
\caption{Nonlinear Hall conductivity(left) and Hall resistivity(right), when $\bm{n}^{R} \parallel [100]$. The Hall conductivity and the Hall resistivity are normalized by the absolute values of the Hall conductivity and the Hall resistivity at $T=0.0001$ for the noninteracting (Free) system, $\sigma_{\mathrm{Hall}}^{0}=0.0198$ and $\rho_{\mathrm{Hall}}^{0}=7.83 \times 10^{-12}$ respectively. Here, we assume a moderate strength of the electric field, $E_{1}=100$ [V/m], which was also assumed in a previous $\it{ab}$ $\it{initio}$ study about the nonlinear Hall effect\cite{Zhang_2018}.}
\label{nonlinear_conductivity_100_Hall}
\end{figure}

Next, let us take a look at the nonlinear Hall conductivity.
When the symmetry reduction appears due to $\bm{n}^{R}  \parallel [111]$, the system's symmetry does not allow for a finite nonlinear Hall conductivity.
However, when the symmetry is reduced via $\bm{n}^{R}  \parallel [100]$, the nonlinear Hall conductivity can be finite.
In particular, $\sigma_{zxx}$ and $\sigma_{zyy}$ belong to the totally symmetric representation of the symmetry $C_{1h}$; thus, they can be finite. In this case, the anti-symmetric part of the nonlinear conductivity tensor with respect to the permutations $(zxx \rightarrow xzx)$ and $(zyy \rightarrow yzy)$ can be finite.
Thus, we define the Hall component of the nonlinear conductivity as
\begin{equation}
\label{hall_definition}
    \sigma_{Hall} = (\sigma_{z11}-\sigma_{1z1})/2
    \end{equation}
We show the temperature dependence of the Hall component of the nonlinear conductivity in Fig.~\ref{nonlinear_conductivity_100_Hall}.
As before, we also include the results for the noninteracting model ("Free") and including only the linear part of the self-energy ("Renormalization").

First, let us take a look at the temperature dependence of the nonlinear Hall conductivity.
As can be seen in Fig.~\ref{nonlinear_conductivity_100_Hall}, the nonlinear Hall conductivity at high temperature is small and negative. As the temperature is lowered, the sign is reversed around the Kondo temperature and the magnitude strongly increases.

The sign change of the nonlinear Hall conductivity is an interesting observation in our results.
Although this behavior strongly depends on the model parameters, we can demonstrate that renormalization effects play a crucial role in the behavior. Here, we consider the nonlinear Hall conductivity at absolute zero and neglect terms from the Fermi sea for simplicity. Thus, we only need to consider contributions from the Fermi surface, enabling us to focus on $\omega=0$.
Then, as can be seen in \eqref{nonlinear_Kubo}, renormalization effects appear only in the $\frac{\partial G^{R}}{\partial \omega}$ which can be written as
\begin{equation}
\label{hall_sign_change}
    \frac{\partial G^{R}}{\partial \omega} = -G^{R} \begin{pmatrix}
1 &
0\\
0 &
1/Z
\end{pmatrix}
G^{R}.
\end{equation}
We can now decompose the Hall conductivity into two parts, the contribution which is proportional to $1/Z$ and the rest, which is not affected by renormalization effects. 
Comparing these two parts for $Z=1$, we see that while the former yields a positive value, the latter yields a negative value.
Moreover, the magnitude of the former is smaller than that of the latter part. Therefore, when $Z=1$, the Hall conductivity is negative, and, when the renormalization effects become strong, the positive contribution overwhelms the negative contribution, which results in a positive nonlinear Hall conductivity. 
Thus, we can observe a sign change when the temperature is lowered. (For the details of the numerical calculations, see appendix \ref{appendix_hall_sign_change}.)
Equation \eqref{hall_sign_change} implies that the positive contribution is connected with the $f$ electrons. Because the $f$ electrons are localized at high temperatures, the dominant contribution to the Hall conductivity at high temperature stems from the itinerant $c$ electrons and the Hall conductivity becomes negative.
In summary, we can say that the renormalization effects drive the sign change.

Next, let us examine how the renormalization effect enhances the magnitude of the nonlinear Hall conductivity.
 In Fig.~\ref{nonlinear_conductivity_100_Hall}, we can see that strong correlations also enhance the nonlinear Hall conductivity compared to the noninteracting system.
Remarkably, while the nonlinear longitudinal conductivity is enhanced $22$ times compared to the free system, the nonlinear Hall conductivity is enhanced by $35$ times. This implies that in the current model the nonlinear Hall effect is enhanced by strong correlations more strongly than other nonlinear conductivities.

In the right panel of Fig.~\ref{nonlinear_conductivity_100_Hall}, we show the Hall resistivity calculated from the linear longitudinal conductivity and the nonlinear Hall conductivity. Generally, the resistivity tensor is defined as the inverse of the conductivity tensor, for example, in a 2D system: $\rho_{xy}=-\sigma_{xy}/(\sigma_{xx}\sigma_{yy}-\sigma_{xy}\sigma_{yx})$. 

In a 3D system, the resistivity tensor becomes more complicated.
However, in the current case, the Hall resistivity can be calculated similarly to the 2D system.
The Hall component originates from the first-order correction to the conductivity by the electric field. Thus, when we focus on the nonlinear Hall effect, we do not have to consider other components of the nonlinear conductivity tensor within the first-order approximation.
Furthermore, the symmetry of the system does not allow for transverse components of the linear conductivity tensor other than the $\sigma_{1z}$.
Thus, we only need to consider the linear conductivity in the $1-z$ plane, and we find for the nonlinear Hall resistivity $\rho_{\mathrm{Hall}} = E_{1} \sigma_{\mathrm{Hall}} / (\sigma_{11}\sigma_{zz}-\sigma_{1z}\sigma_{z1})$.

As is shown in Fig.~\ref{linear_conductivity} (and Fig.~\ref{linear_conductivity_100} in appendix \ref{appendix_dmft}), the longitudinal conductivity $\sigma_{11}$
monotonically decreases as the temperature is lowered. Also, $\sigma_{zz}$ and $\sigma_{1z}$ behave almost the same as $\sigma_{11}$ except that the magnitude of $\sigma_{1z}$ is much smaller than that of $\sigma_{zz}$ and $\sigma_{11}$. Furthermore, as is shown in Fig.~\ref{nonlinear_conductivity_100_Hall}, the magnitude of the nonlinear Hall conductivity $\sigma_{Hall}$ 
increases as the temperature is lowered. Thus, the qualitative temperature dependence of the Hall resistivity is almost the same as that of the nonlinear Hall conductivity. Besides, the enhancement of the magnitude of the Hall resistivity directly corresponds to that of the nonlinear Hall conductivity because the linear longitudinal conductivity is not enhanced by  renormalization effects at low temperature. On the other hand, at high temperature, the Hall resistivity is suppressed by the large linear longitudinal conductivity.

\section{Comparison to Experiment}
\subsection{Longitudinal Resistivity}
We now want to directly compare our results with the experimental observations in reference \cite{Dzsaber_2021,dzsaber_weyl_kondo_2017}.
The characteristic behavior of the linear resistivity found in the experiment is that the resistivity increases when lowering the temperature below the Kondo temperature\cite{dzsaber_weyl_kondo_2017,Dzsaber_2021}.

First, we see that there are only tiny quantitative differences comparing our results of the linear longitudinal conductivity between the systems with and without Rashba spin-orbit coupling. Thus, when comparing our calculations of the linear conductivity or resistivity to the experiments, we do not need to consider whether or how the symmetry of the system is reduced. 
By analyzing spectral functions, we found that the Kondo temperature in our model is about $T_{\mathrm{K}} = 0.0016$. Lowering the temperature below this temperature, we see that the resistivity increases before becoming constant at low temperatures.
This behavior is in good agreement with the experimental observation and arises due to the formation of a Weyl semimetal at low temperatures.
We note that when we use the Kubo formula to calculate conductivities numerically, we need to introduce the inverse of an impurity scattering lifetime ($\delta$) to the denominator of the Green's function. We here take a relatively large value ($\delta=0.1$) to reduce the computation cost to calculate the integrals in the Kubo formula.
Due to this scattering term, the resistivity saturates at low temperatures.
When using a smaller $\delta$, we expect that the resistivity increases more sharply around the Kondo temperature as in the experiments\cite{dzsaber_weyl_kondo_2017,Dzsaber_2021}.

\subsection{Hall Resistivity}
In our model, the symmetry of the Hamiltonian must be reduced from the original symmetry group of $\mathrm{Ce_{3}Bi_{4}Pd_{3}}$ so that a finite nonlinear Hall effect can be observed. Therefore, when comparing with the experiment, we only focus on the calculations, including the Rashba spin-orbit coupling with $\bm{n}^{R} \parallel [100]$.

The important features observed in the experiment on $\mathrm{Ce_{3}Bi_{4}Pd_{3}}$ are that the nonlinear Hall effect is observed only below the Kondo temperature and that the Hall resistivity increases as the temperature is lowered. 
Furthermore, the Hall angle of $\mathrm{Ce_{3}Bi_{4}Pd_{3}}$ is $10^{3}$ times as large as that of weakly interacting Weyl semimetals.

In our calculations, the Hall resistivity is small at high temperatures and starts to increase when lowering the temperature below the Kondo temperature.
Thus, the behavior that the Hall resistivity arises below the Kondo temperature is in agreement with the experiment.
The sign change of the Hall resistivity found in our calculations has not been observed in the experiment.
However, as we have noted above, this behavior strongly depends on the parameters. Furthermore, as the Hall resistivity was too small to be observable in the experiments at high temperatures, a possible sign change could not be detected. 

Finally, a crucial feature of the experiment is the magnitude of the Hall resistivity. Our calculations have demonstrated that the renormalization effect strongly enhances nonlinear conductivities. A heat capacity experiment has revealed that the mass of the electrons is strongly enhanced by about $10^{3}$ times in $\cebipd$\cite{dzsaber_weyl_kondo_2017}. Thus, following our considerations, the nonlinear Hall effect should be enhanced by $10^{3}$ times, which is consistent with the experimental fact that the (nonlinear) Hall angle of $\cebipd$ is $10^{3}$ times larger than that of a weakly interacting Weyl semimetal.
Therefore, the enhancement of nonlinear responses by strong correlations, which is also found in our calculations, is most likely the origin of the giant spontaneous Hall effect observed in $\cebipd$.

\section{Summary}

We have studied strong correlation effects on nonlinear responses by using DMFT focusing on the nonlinear longitudinal and Hall conductivities in a Weyl-Kondo semimetal. 
Our calculations have revealed that the longitudinal resistivity of the Weyl-Kondo semimetal strongly increases when lowering the temperature below the Kondo temperature. Furthermore, we have shown that the Hall resistivity also strongly increases below the Kondo temperature if the symmetry of the system allows for a nonlinear Hall effect. Both results are consistent with the experiments on $\cebipd$. Remarkably, our numerical calculations have shown that strong correlations enhance the nonlinear Hall effect. Thus, our calculations give a possible theoretical explanation of the giant nonlinear Hall effect in $\cebipd$.

In particular, our analyses have shown that second-order nonlinear conductivities, e.g., the nonlinear nonreciprocal conductivity, are enhanced by strong correlations due to the renormalization of the band structure.
Thus, a combination of two effects, the decrease of the linear longitudinal conductivity due to the formation of a Weyl-semimetal and the enhancement of the nonlinear longitudinal conductivity due to correlations, makes it easy to observe a large nonreciprocal response at low temperature in this kind of materials.
Accordingly, the enhancement of nonlinear responses due to strong correlations implies that strongly correlated electron systems, including Weyl-Kondo semimetals, are promising platforms for realizing giant nonlinear responses. Finally, the sign change in the Hall conductivity and the Hall resistivity is an interesting observation in our calculations, which might be observed experimentally in the future, and provides valuable information about the detailed band structure.

\section*{Acknowledgements}
We thank Hikaru Watanabe, Riki Toshio, Youichi Yanase for insightful discussions.
R.P. is supported by JSPS, KAKENHI Grant No. JP18K03511. Y. M. is supported by WISE program, MEXT, JSPS reserch fellowship, and by JSPS KAKENHI Grant No. 20J12265.
The computation in this work has been done using the facilities of the Supercomputer Center, the Institute for Solid State Physics, the University of Tokyo.

\appendix

\section{Spectral functions \& Linear Resistivity}
\label{appendix_dmft}
In this appendix, we show the self-energies, the spectral functions, and the linear resistivity in the system with further symmetry reduction($\bm{n}^{R}\parallel[111]$ and $\bm{n}^{R}\parallel[100]$).
As we have noted in the first paragraph of section \ref{E_is_considered}, the symmetry reduction does not strongly affect these physical quantities.

\begin{figure*}[t]
\begin{center}
\includegraphics[width=0.32\linewidth]{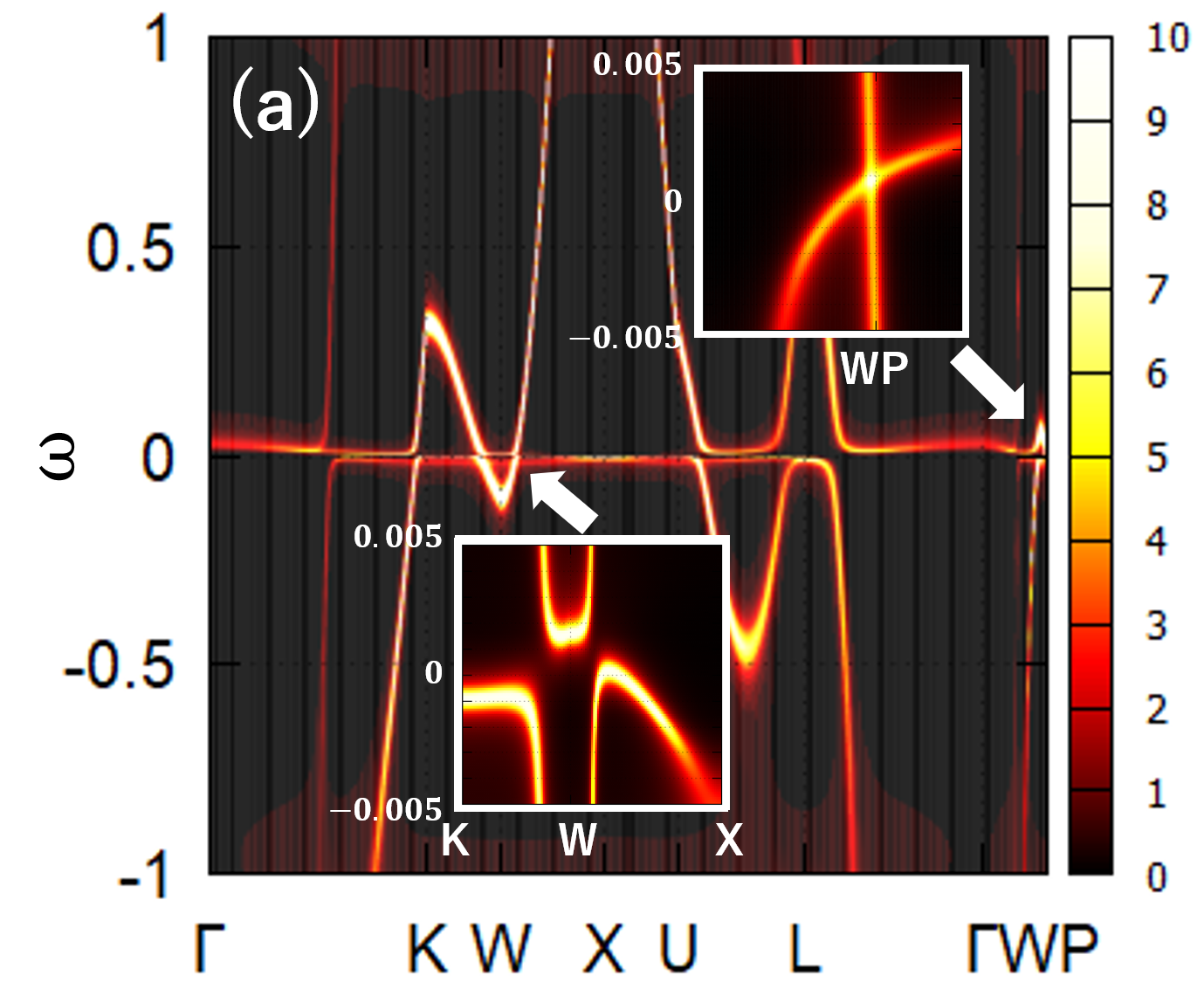}
\includegraphics[width=0.32\linewidth]{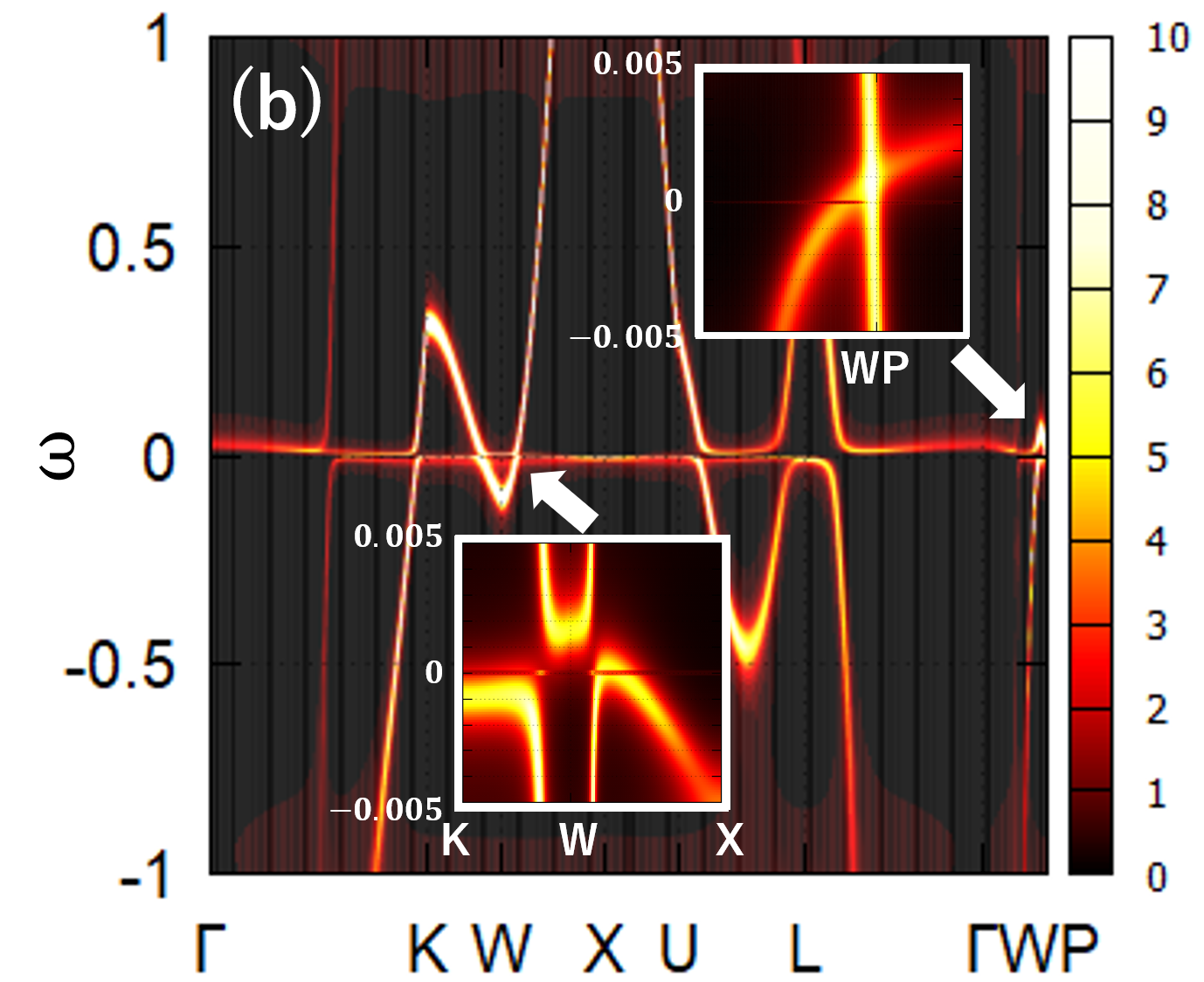}
\includegraphics[width=0.32\linewidth]{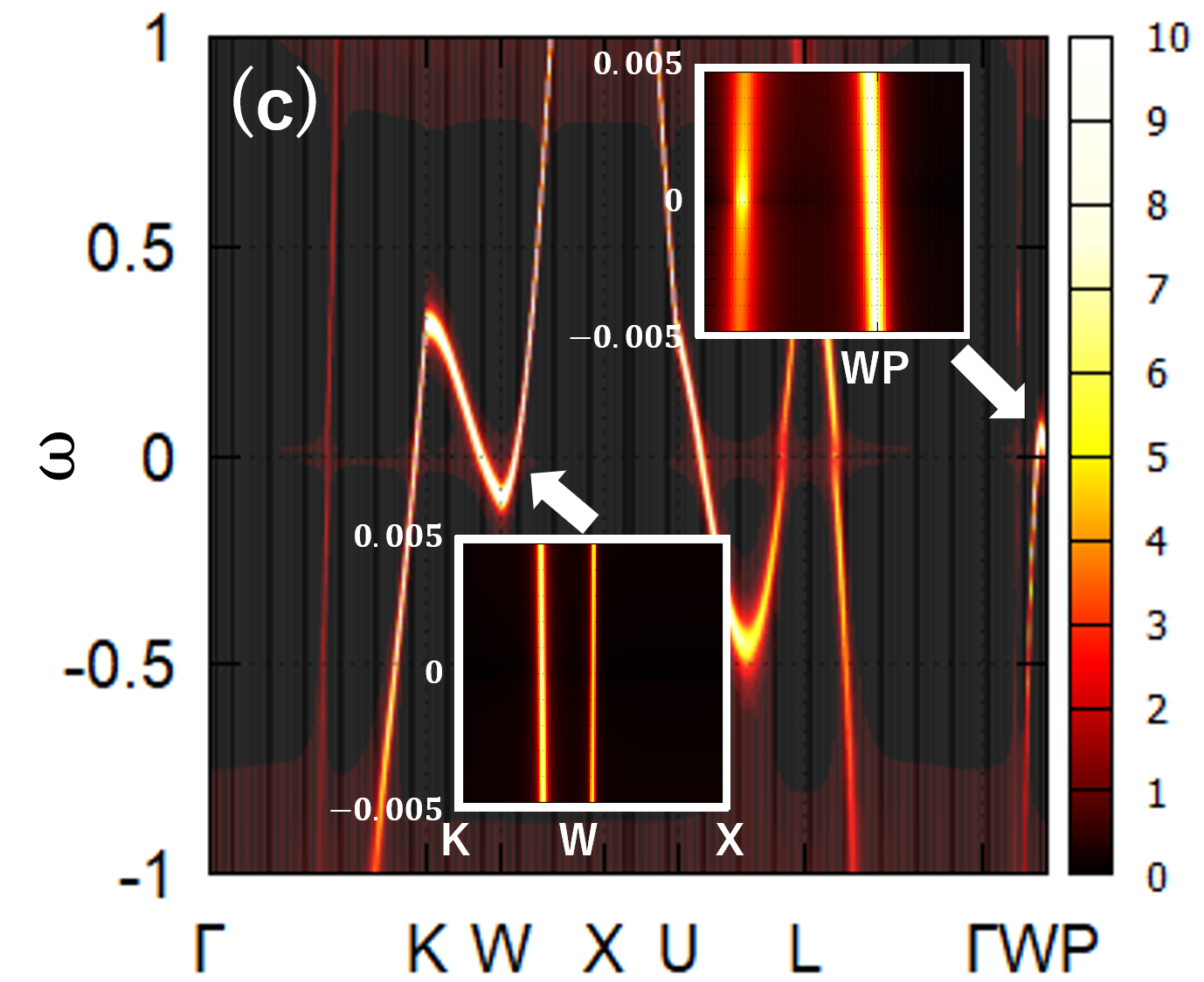}
\end{center}
\caption{Spectral functions (from left to right, $T=0.0001$, $T=0.0016$ and $T=0.0256$) when $\bm{n}^{R} \parallel [100]$. WP denotes the momentum where one of the Weyl points lies at:(3.02,1.43,0.120).}
\label{spectrum_111}
\end{figure*}

\begin{figure*}[t]
\begin{center}
\includegraphics[width=0.32\linewidth]{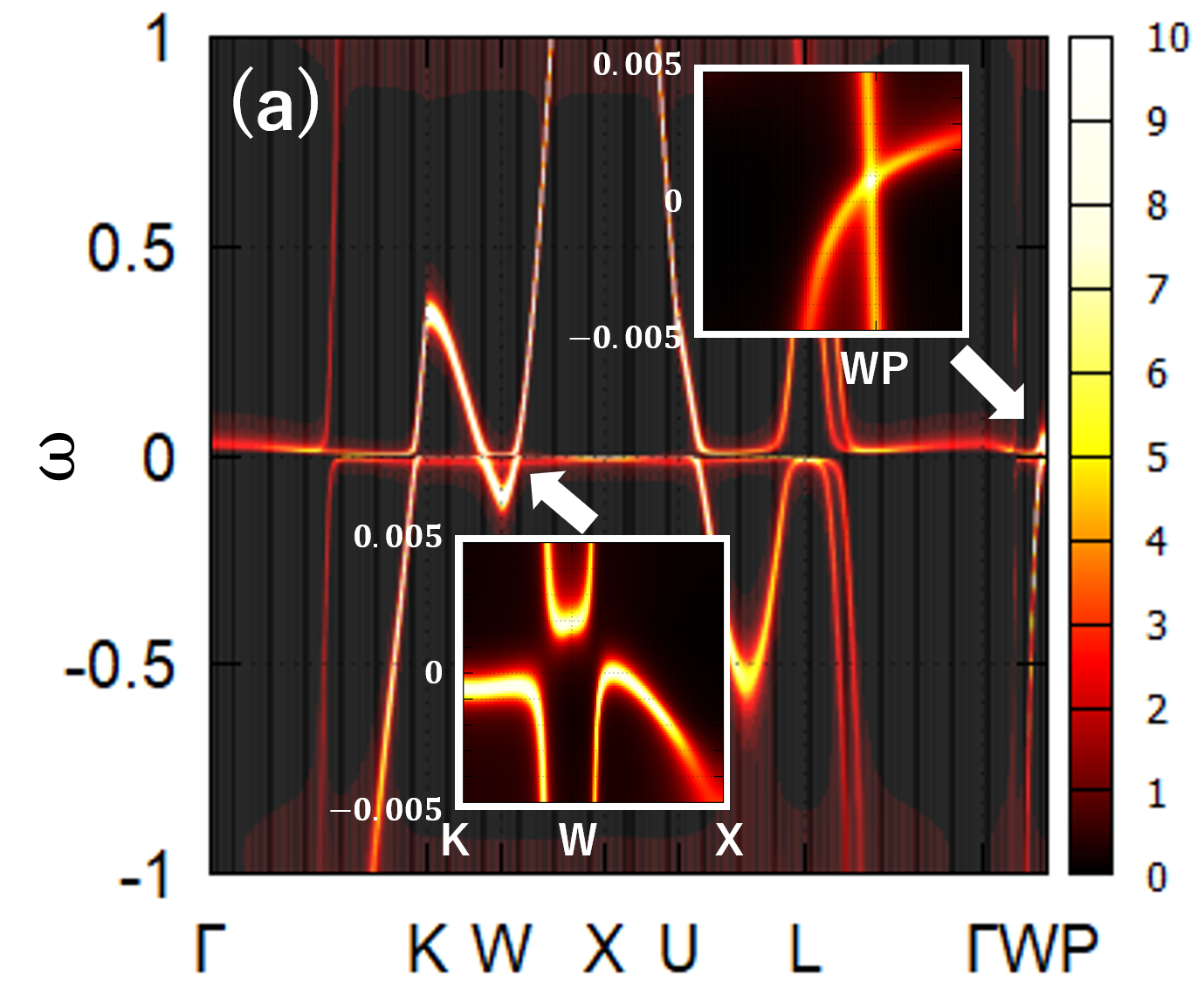}
\includegraphics[width=0.32\linewidth]{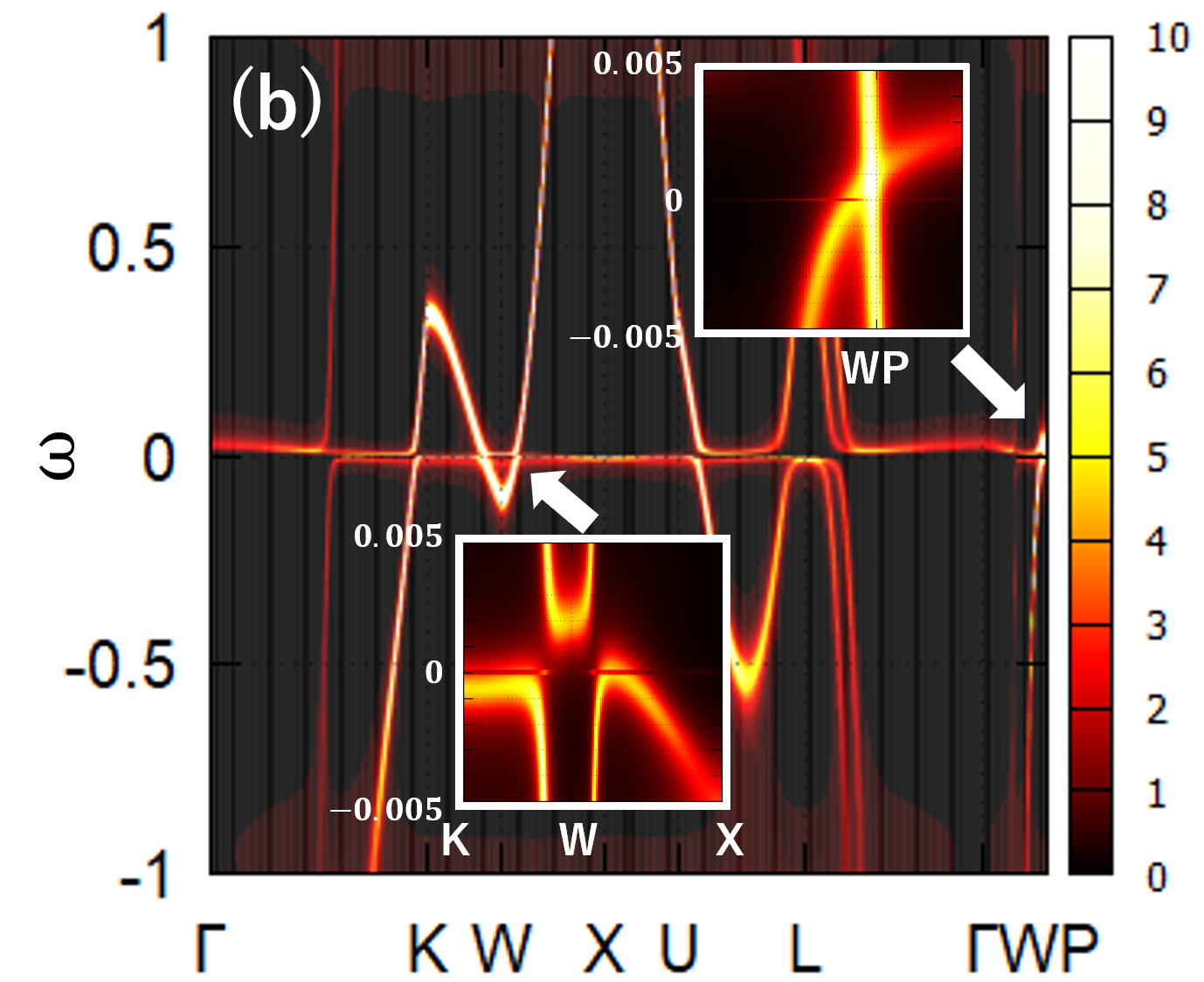}
\includegraphics[width=0.32\linewidth]{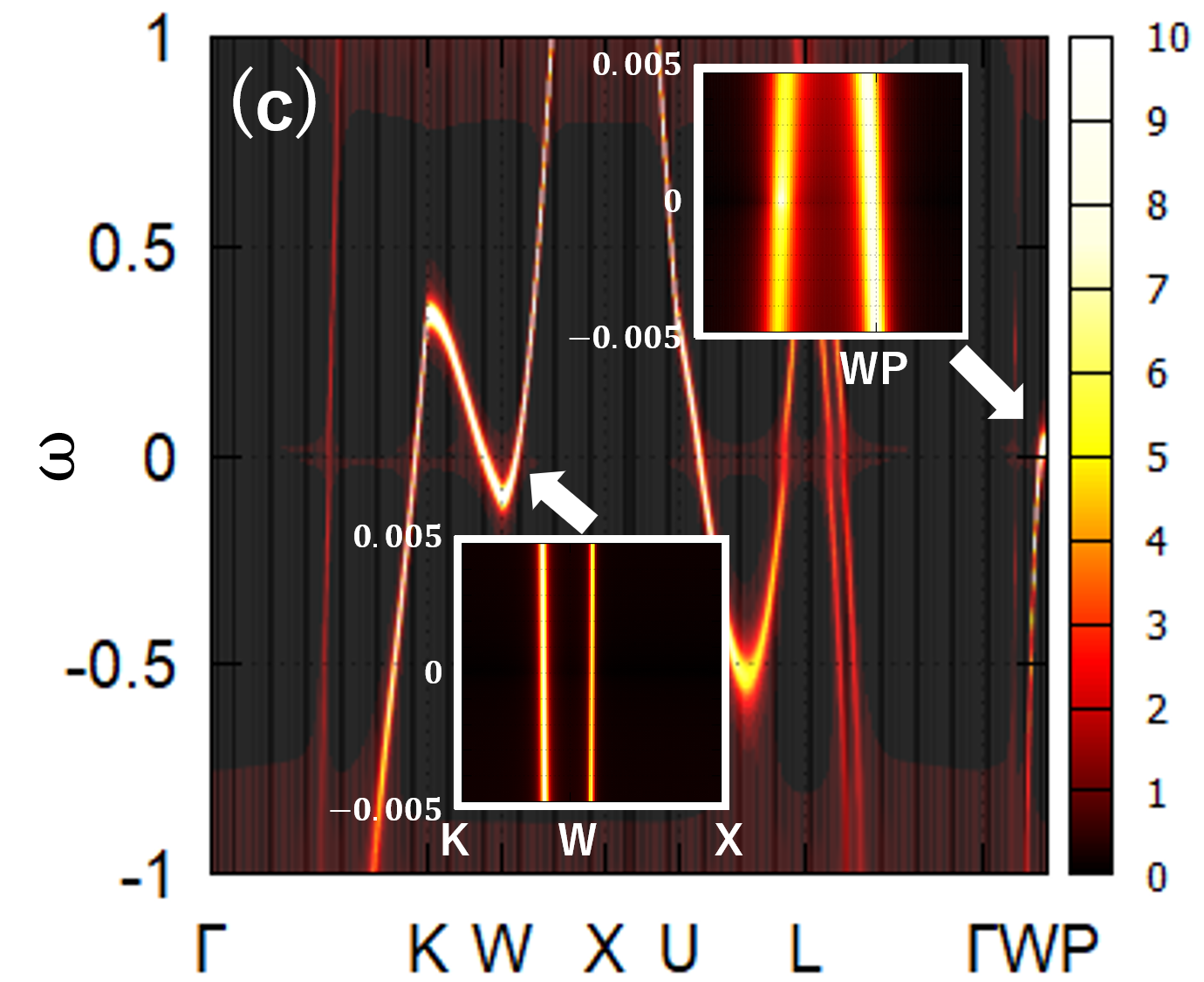}
\end{center}
\caption{Spectral functions (from left to right, $T=0.0001$, $T=0.0016$ and $T=0.0256$) when $\bm{n}^{R}  \parallel [111]$. WP denotes the momentum where one of the Weyl points lies at:(3.13,1.38,0.140).}
\label{spectrum_100}
\end{figure*}

\begin{figure}[t]
\begin{center}
\includegraphics[width=0.49\linewidth]{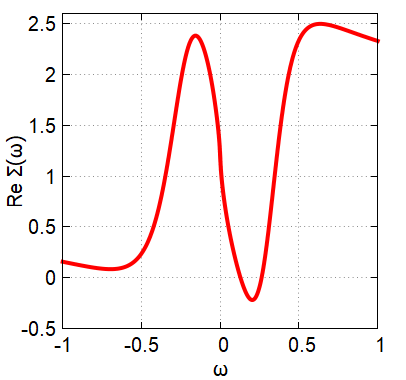}
\includegraphics[width=0.49\linewidth]{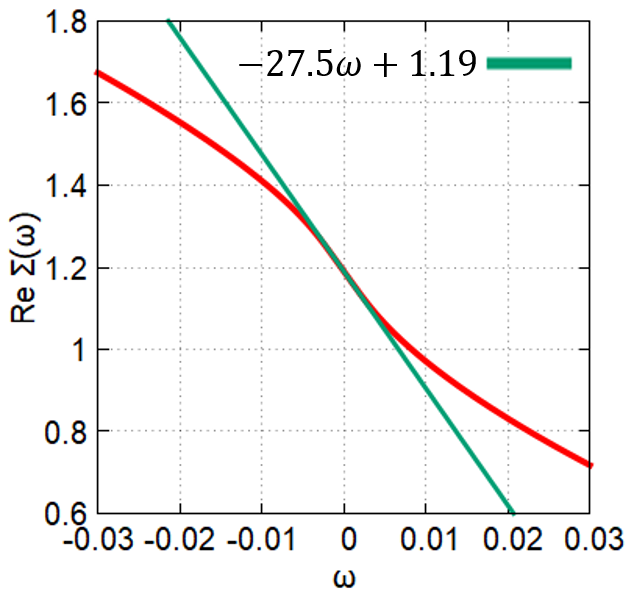}
\end{center}
\caption{Left: Real part of the self-energy when $\bm{n}^{R}  \parallel [111]$ for $T=0.0001$. Right: Magnification around the Fermi energy with linear fit.}
\label{selfenergy_111}
\end{figure}

\begin{figure}[t]
\begin{center}
\includegraphics[width=0.49\linewidth]{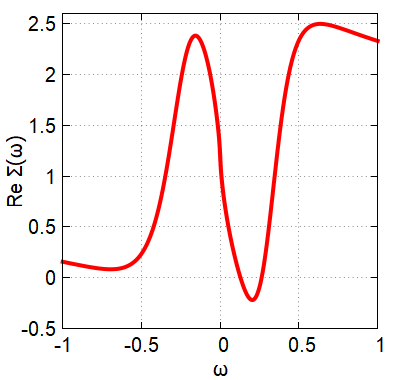}
\includegraphics[width=0.49\linewidth]{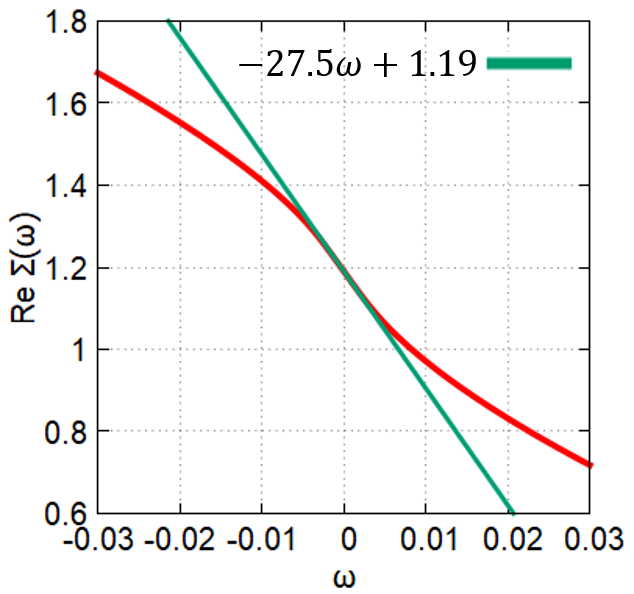}
\end{center}
\caption{Same as Fig. \ref{selfenergy_111} for $\bm{n}^{R} \parallel [100]$}
\label{selfenergy_100}
\end{figure}

In Figs. \ref{spectrum_111} and \ref{spectrum_100}, we plot the spectral functions between the high-symmetry points and the Weyl point, denoted by WP, for three characteristic temperatures, $T=0.0001$, $T=0.0016$, $T=0.0256$ (from the left to right) when $\bm{n}^{R}\parallel[111]$ and $\bm{n}^{R}\parallel[100]$ respectively.
The parameters of the Rashba spin-orbit coupling are as follows: $\lambda_{c}^{R} = -0.32$, $\lambda_{f}^{R} = 0.016$, and $\lambda_{cf}^{R}=-0.04$.
These figures show that the Weyl points between the $W$ point and $X$ point gap out. However, six Weyl points remain after the symmetry reduction which is shown in the insets at the top of the Figs. \ref{spectrum_111} and \ref{spectrum_100}.
The corresponding real parts of the self-energies at $T=0.0001$ are shown in Fig.~\ref{selfenergy_111} and \ref{selfenergy_100} respectively.
As can be seen in the right panel of Figs.~\ref{selfenergy_111} and \ref{selfenergy_100}, in both cases, the slope of the real part of the self-energy at $\omega=0$ is about $27.5$. Thus, the mass of an electron in the vicinity of the Fermi surface is enhanced about $28.5$ times.
The renormalization effects in these systems are slightly smaller than that in the system without further symmetry reduction.
Also, based on calculations of the spectral functions for different temperatures, we estimate the Kondo temperatures in these cases to $T_{\mathrm{K}} \sim 0.0016$.

\begin{figure}[t]
\begin{center}
\includegraphics[width=0.49\linewidth]{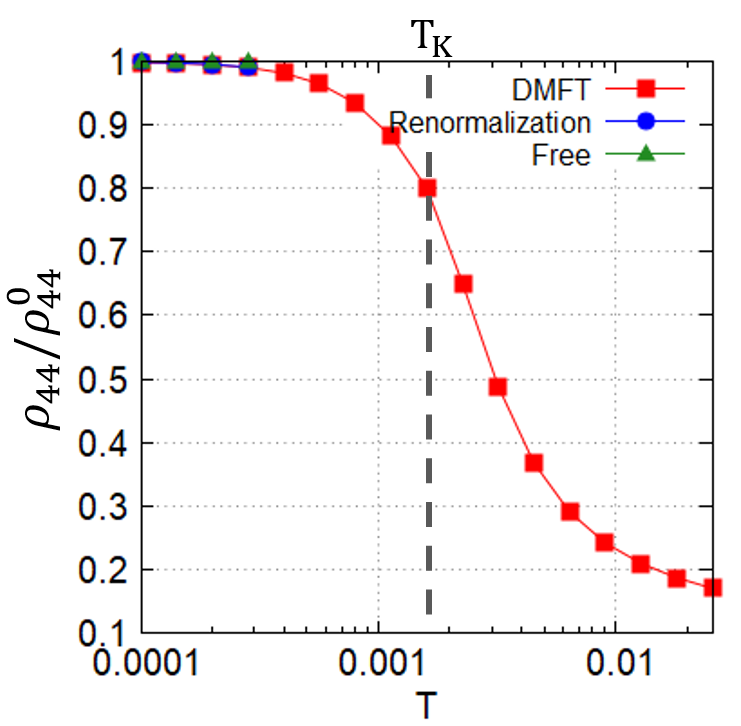} 
\includegraphics[width=0.49\linewidth]{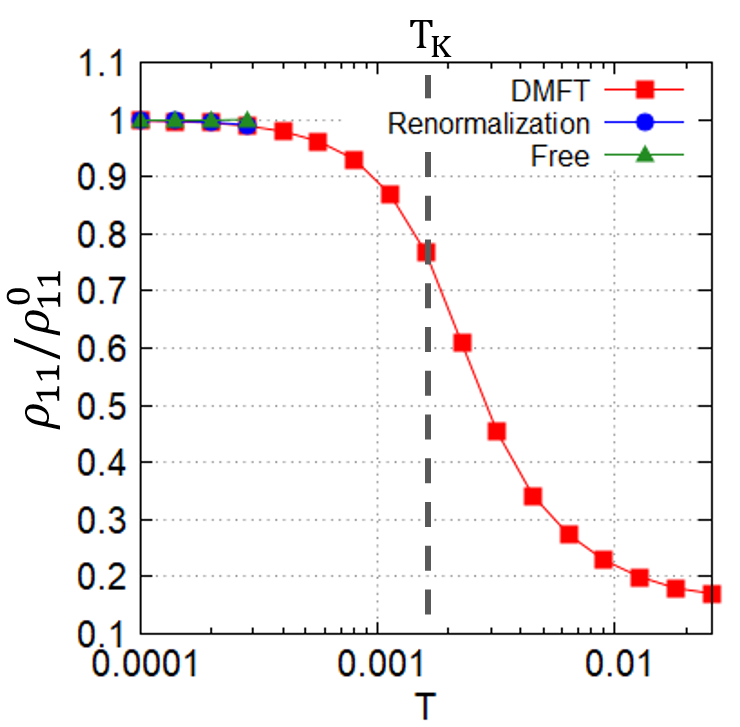}
\end{center}
\caption{Linear longitudinal resistivity; The left (right) panel shows the resistivity when $\bm{n}^{R}\parallel[111]$ ($\bm{n}^{R}\parallel[100]$). The resistivities are normalized by the corresponding resistivities for  $T=0.0001$ of the noninteracting (Free) system, $\rho_{44}^{0}=1.75$ and $\rho_{11}^{0}=1.98$ respectively.}
\label{linear_conductivity_100}
\end{figure}

In Fig.~\ref{linear_conductivity_100}, we show the longitudinal resistivity in the systems with further symmetry reduction when $\bm{n}^{R}\parallel[111]$ (left panel) and $\bm{n}^{R}\parallel[100]$ (right panel) respectively.
These figures show that, in both cases, the resistivity increases as the temperature is lowered below the Kondo temperature and  renormalization effects do not affect the magnitude of the resistivity at low temperatures.
This behavior is the same as in the system without further symmetry reduction.

\section{Sign change of Hall conductivity}
\label{appendix_hall_sign_change}
\begin{figure}[t]
\begin{center}
\includegraphics[width=0.99\linewidth]{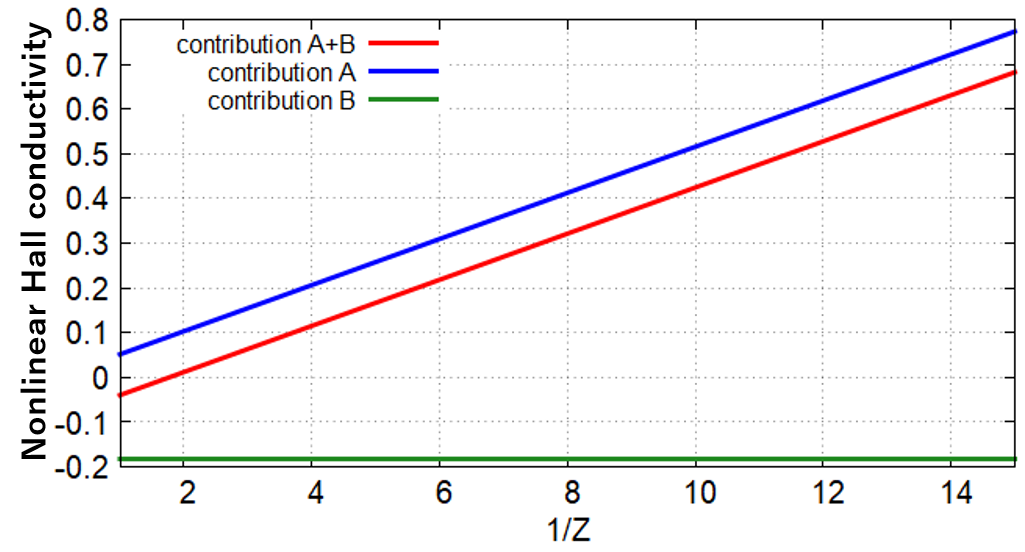}
\end{center}
\caption{$1/Z$ dependence of the renormalized and unrenormalized contribution to the nonlinear Hall conductivity. Contribution A is the term that is proportional to $1/Z$, and contribution B is independent of $1/Z$.  A+B corresponds to the full nonlinear Hall conductivity.}
\label{nonlinear_hall_sign_change}
\end{figure}

As explained above, the nonlinear Hall conductivity at zero temperature can be written as a sum of a term affected by the renormalization and a term independent of the renormalization.
Here, we analyze these two different contributions to the nonlinear Hall conductivity.
In Fig.~\ref{nonlinear_hall_sign_change}, we show the $1/Z$ dependences of the former contribution, denoted by "contribution A", the latter contribution, denoted by "contribution B," and the sum of these contributions, denoted by "contribution A+B".
As can be seen in Fig.~\ref{nonlinear_hall_sign_change}, contribution A always gives a positive contribution to the nonlinear Hall conductivity. In contrast, contribution B is always negative. 
At weak renormalization strength, $Z \sim 1$, the magnitude of  contribution B is larger than that of contribution A.
Thus, the nonlinear Hall conductivity is negative at weak renormalization.
On the other hand, contribution A grows linearly as the renormalization increases. At some critical renormalization, contribution A overwhelms  contribution B, and the nonlinear Hall conductivity becomes positive.

In our calculations in section \ref{E_is_considered}, $1/Z$ is about $28.5$, which is strong enough to change the sign of the nonlinear Hall conductivity in the interacting case compared to that in the noninteracting case.

\bibliography{bunken}

\end{document}